\documentclass[twocolumn]{aastex631}
\usepackage{booktabs}
\usepackage{natbib}
\usepackage{booktabs}
\usepackage{amsmath}
\usepackage{hyperref}
\usepackage{cleveref}
\usepackage[section]{placeins}

\shorttitle{Aferglows Of Fast Radio Bursts}
\shortauthors{Bian \& Deng}
\graphicspath{{./}{Images/}}

\begin{document}
	
	\title{Prediction of Multi-Wavelength Afterglows Associated with FRB 20200120E and FRB 20201124A}
	\author{Ke Bian}
	
	\author[0000-0003-0471-365X]{Can-Min Deng}

	\affiliation{
		Guangxi Key Laboratory for Relativistic Astrophysics, Department of Physics, Guangxi University, Nanning 530004, China; dengcm@gxu.edu.cn}

\begin{abstract} 
Fast radio bursts (FRBs) are mysterious radio transients with uncertain origins and environments. Recent studies suggest that some active FRBs may originate from compact objects in binary systems. In this work, we develop a unified theoretical framework to model the multi-wavelength afterglows of FRBs resided in binary systems and apply it to two representative repeaters, FRB 20200120E and FRB 20201124A. By solving the dynamics and radiation processes of FRB ejecta interacting with the surrounding medium, we compute afterglow light curves in the radio, optical, and X-ray bands. Our results show that radio afterglows offer the best prospects for detection, with their brightness highly sensitive to ejecta kinetic energy and ambient density. Future high-sensitivity radio telescopes, such as the Square Kilometre Array (SKA), could detect these signals. Optical afterglows, though short-lived and challenging to observe, may be significantly enhanced in dense environments, potentially making them detectable with facilities like the Large Synoptic Survey Telescope (LSST). In contrast, X-ray afterglows are predicted to be too faint for detection with current instruments. Our study highlights the potential of multi-wavelength afterglows as probes of FRB progenitors and their surrounding environments, offering crucial insights into the nature of these mysterious transients.

	\end{abstract}

\section{Introduction} \label{sec}
Fast Radio Bursts (FRBs) are millisecond-duration radio transients of extragalactic origin, first discovered by \citep{2007Sci...318..777L,2013Sci...341...53T}. Over the past decade, hundreds of FRBs have been detected, with some sources exhibiting repetition while others appear to be one-off events  \citep{2023Univ....9..330X}. Despite extensive observational efforts, the physical origins of FRBs remain uncertain. Many theoretical models have been proposed to explain their production mechanisms \citep{2019ARA&A..57..417C,2019A&ARv..27....4P,2020Natur.587...45Z,2021SCPMA..6449501X,2022A&ARv..30....2P,2023RvMP...95c5005Z,2024Ap&SS.369...59L}. 
%which can be broadly classified into magnetar-based models, binary interaction models, and accretion-driven models.
One of the most widely discussed explanations for FRBs is that they originate from magnetars, where bursts are powered by the sudden release of magnetic energy \citep{2014MNRAS.442L...9L,Beloborodov+2017,Metzger+2019, Beloborodov+2020, Lu+2020}. This model naturally explains the association between FRB 20200428D and the Galactic magnetar SGR 1935+2154, which provided the first direct evidence linking FRBs to magnetars \citep{2020Natur.587...54C,2020Natur.587...59B}. 

While magnetars provide a compelling explanation for some FRBs, they do not fully account for all observed FRB properties. For instance, the long periodic activity seen in FRB 20180916B and FRB 20121102A is difficult to reconcile with an isolated magnetar, leading to the exploration of alternative models, particularly those involving binary systems \citep{2020ApJ...890L..24Z,2020ApJ...893L..39L,2020ApJ...893L..26I,2020MNRAS.497.1543G,2020A&A...644A.145M,2020ApJ...895L...1D,2020ApJ...893L..31Y,Deng_2021,2021ApJ...917...13S,2021ApJ...918L...5L}. 
In addition to periodic activity, recent observations have revealed that some FRBs exhibit erratic variations in their Faraday rotation measures (RMs), including large magnitude changes and even sign reversals. For instance, FRB 20190520B and FRB 20201124A show complex RM variations, which are thought to arise from interactions between the FRB source and a dense plasma environment \citep{2022NatCo..13.4382W, 2023Sci...380..599A}. These observations suggest that some FRBs may originate from binary systems where the surrounding plasma is influenced by the companion’s stellar wind or decretion disk. \citet{2022NatCo..13.4382W} proposed that FRB 20201124A and FRB 20190520B may be produced in magnetar/Be star binary systems. Additionally, \citet{2023ApJ...942..102Z} and \citet{2023A&A...673A.136R} suggested that the RM variations in repeating FRBs could be explained by binary systems, where interactions with the companion’s stellar wind or outflows lead to time-dependent magneto-ionic environments. These findings further support the idea that some FRBs are produced in binary systems, with their environments playing a crucial role in shaping their observed properties.

{Similar to gamma-ray bursts (GRBs), FRB explosions may also eject material, which then interacts with the surrounding environment to produce multi-wavelength afterglow emission \citep{2014ApJ...792L..21Y}.
Theoretically, FRB models involving neutron stars can be broadly divided into two categories. The first involves inner magnetospheric models, where coherent emission originates near the neutron star surface, such as through magnetic reconnection or plasma instabilities \citep[e.g.,][]{2020MNRAS.494.2385K,Lu+2020,2020ApJ...897....1L,2024ApJ...972..124Q}. These models generally do not predict detectable afterglow emission. The second class comprises baryonic outflow models, in which relativistic ejecta launched during the burst interact with the surrounding medium, leading to synchrotron afterglows in radio, optical, or X-ray bands \citep[e.g.,][]{2014MNRAS.442L...9L,Beloborodov+2017,Metzger+2019,Beloborodov+2020,Deng_2021,2021ApJ...917...13S}. Afterglow observations can thus help distinguish between these two classes of models.
The nature of the surrounding environment depends on the specific FRB progenitor scenario, leading to different predicted afterglow behaviors and evolution patterns \citep{2018pgrb.book.....Z}. Therefore, studying FRB afterglows is not only intrinsically interesting but also provides a potential means to test and constrain competing FRB models.}
%Similar to gamma-ray bursts (GRBs), FRB explosions may also eject material, which then interacts with the surrounding environment to produce multi-wavelength afterglow emission \citep{2014ApJ...792L..21Y}. The nature of the surrounding environment depends on the specific FRB models, leading to different predicted afterglow behaviors and evolution patterns \citep{GAO2013141}. Therefore, studying FRB afterglows is not only intrinsically interesting but also provides a potential means to distinguish between various FRB models.
For cosmological FRBs, detecting their afterglows is particularly challenging due to their immense distances and the faintness of the resulting afterglows, which often fall below the sensitivity limits of current telescopes \citep{2021Univ....7...76N}. However, nearby FRBs present a promising opportunity for afterglow searches. A notable example is FRB 20200428, which originated from a Galactic magnetar \citep{2020Natur.587...54C,2020Natur.587...59B}.
\citet{2022MNRAS.517.5483C} have modeled the afterglow of FRB 20200428 based on the flaring magnetar model proposed by \citet{Metzger+2019}, identifying observable afterglow signatures under specific conditions. Additionally, \citet{2023RAA....23k5010D} examined the afterglow properties of FRB 20200428 in a model-independent way. Although this FRB was much lower in energy than extragalactic bursts, its proximity makes its afterglow potentially detectable. However, its extremely low repetition rate has limited the potential for follow-up observations aimed at detecting the afterglow.

Notably, \citet{Bhardwaj_2021,2022Natur.602..585K} reported that FRB 20200120E is an active repeating source located in the globular cluster of M81. Recent observations have detected a burst from FRB 20200120E with a fluence of 30 Jy ms, corresponding to an energy of approximately \(10^{36}\) erg \citep{2024NatCo..15.7454Z}. This energy level is comparable to the low-energy bursts of FRB 20121102A \citep{2021Natur.598..267L}. Moreover, its proximity at only 3.6 Mpc makes it a promising candidate for detecting potential afterglow emission. Given its distinct environment within a globular cluster and its observed burst properties, FRB 20200120E offers a unique opportunity to investigate possible afterglow signatures in a different astrophysical setting.

FRB 20201124A is an active FRBs located in a barred spiral galaxy, and its repetition rate can reach up to 545 per day \citep{2022Natur.609..685X}. This FRB is not only highly active but also very bright. According to \cite{2022MNRAS.512.3400K}, the strongest burst from this source reached a fluence density of 640 Jyms, setting a new brightness record for extragalactic FRBs.
The isotropic-equivalent energy of this burst is estimated to be $\sim$ \(10^{41}\) erg. Given its distance of 453 Mpc \citep{2022Natur.609..685X}, the high energy release suggests that its potential afterglow could still be detectable even at such a distance.

Inspired by that, in this paper, we model the possible multi-band afterglow and its evolution associated with FRB 20200120E and FRB 20201124A, and discussing its detectability. By comparing the theoretical predictions with possible observational constraints in the further, it will help to reveal the formation mechanisms of FRBs and the properties of their surrounding environments, ultimately contributing to a better understanding of the physical origins of FRBs.

This paper is organized as follows. In Section 2, we describe our theoretical framework. Section 3 presents our predicted afterglow light curves. Section 4 discusses observational implications and detection strategies. Finally, Section 5 summarizes our conclusions.

\section{THE MODEL} \label{Sec.2}
In this section, we describe the details of the model and methods used to calculate multiwavelength afterglow emission. 
FRB 20200120E is located within a globular cluster of the M81 galaxy, leading to speculation that it resides in the most common system found in such environments—a low-mass accreting binary \citep{2023Natur.618..484L}. Given that neutron stars in LMXBs typically possess significantly weaker magnetic fields compared to magnetars, this suggests that FRB 20200120E is more likely powered by an accreting neutron star (or black hole) rather than a magnetar \citep{Deng_2021,2021ApJ...917...13S}.  
Interestingly, FRB 20201124A has also been proposed to reside in a binary system, with its companion being a massive Be star, which can naturally explain its features like RM variation \citep{2022NatCo..13.4382W}.  In this scenario, FRB 20201124A could also be generated by a neutron star accreting material from a Be companion \citep{Deng_2021}.
Therefore, in the following, we present the theoretical framework for the afterglow emission associated with FRB 20200120E and FRB 20201124A  within the context of the FRB model proposed by \citet{Deng_2021}.

%The key parameters of our model include the total kinetic energy \(E\), the Lorentz factor \(\eta\), the shock energy equipartition parameters \(\epsilon_e\) and \(\epsilon_B\) for electrons and magnetic fields, the electron injection spectral index \(p\), and the number density of the ambient medium \(n\).
\subsection{Dynamics}
According to this model, FRBs are generated by the interaction between relativistic ejecta from an accreting neutron star and the surrounding medium. Following the FRB emission, the relativistic ejecta\footnote{ {In this study, we assume a spherically symmetric outflow for the relativistic ejecta. This assumption is motivated by the lack of direct observational constraints on the outflow geometry in FRBs. While GRB afterglows are known to arise from collimated jets with finite opening angles, it remains uncertain whether similar jet structures exist in FRBs, especially given their potentially different progenitor environments and energetics. Should future observations suggest anisotropic or collimated outflows in FRBs, our framework could be generalized to include jet dynamics and jet-break effects. Nevertheless, for the purpose of exploring detectability of afterglow emission in a model-independent way, the spherical case provides a conservative and tractable baseline.}} continues to propagate outward, sweeping up ambient material and producing synchrotron afterglow emission, analogous to afterglows observed in gamma-ray bursts (GRBs). We consider this relativistic ejecta with an initial kinetic energy $E_k$ and an initial Lorentz factor $\eta$.  As the ejecta propagates into the surrounding medium, it undergoes deceleration due to the accumulation of swept-up material.

{In this work, we adopt a theoretical framework inspired by the well-established GRB afterglow formalism to model potential multi-wavelength afterglows associated with FRBs. While we acknowledge that FRBs and GRBs differ in several respects, such as progenitor systems, outflow composition, and energetics, the absence of confirmed FRB afterglow detections and the current uncertainties in FRB progenitor models motivate a simplified, phenomenological approach. In fact, most existing studies exploring FRB afterglow emission (e.g., \citealt{2014ApJ...792L..21Y,2020ApJ...899L..27M,2022MNRAS.517.5483C,2023RAA....23k5010D}) adopt a similar treatment, emphasizing order-of-magnitude estimates and detectability forecasts rather than precise fits.
Our primary objective is to assess the detectability of afterglow signals under physically motivated assumptions, thereby providing guidance for future observations. This modeling philosophy is analogous to the early development of GRB afterglow theory, which relied on idealized parameter studies before observational constraints were available \citep{Sari_1998}. Given the lack of consensus on key microphysical parameters, such as the outflow geometry, magnetization, and baryon loading, we opt for a transparent parametrization of the external shock evolution, focusing on key factors such as kinetic energy and ambient density. We note that detailed microphysics, such as radiative efficiencies and energy partition parameters, introduce uncertainties that may be comparable to or larger than those due to simplifying assumptions in geometry or outflow structure. 
This approach allows us to systematically explore a range of plausible scenarios while remaining agnostic about the specific FRB central engine.}

A distinctive feature of this model is the special peristellar environment, which is filled with material from the companion star's stellar wind and extends into the interstellar medium. 
For the binary system, the orbital separation is determined by the Kepler's third law
\begin{eqnarray}
a \simeq \left(\frac{4\pi^2T^2}{GM_{\rm tot}}\right)^{1/3}
\end{eqnarray}
where  \(G\) is the  gravitational constant, \(T\) is the orbital period,  and $M_{\rm{tot}}=M_{\rm{NS}}+M_{\rm{C}}$ is the total mass of the binary. $M_{\rm{NS}}=1.4M_{\odot}$ is the mass of the neutron star. $M_{\rm{C}}$ denotes the mass of the companion star, with its specific value determined by the detailed research content discussed subsequently.

We introduce a parameter that is the number density $n_a$ of the stellar wind from the companion star at a distance of $a$ from it:
\begin{eqnarray}
n_a =\frac{\dot{M}_{\rm w}}{4 \pi m_{\rm p} a^2 v_{\rm w}},
\end{eqnarray}
where \(\dot{M}_{\rm w}\) is the wind mass-loss rate in units of solar masses per year and \(v_{\rm w}\) is the wind velocity. Thus, at a distance r from the neutron star in the direction perpendicular to the orbital plane, the number density of the surrounding medium  can be expressed as \citep{Deng_2021}
\begin{eqnarray}
n(r) =a^2n_a/(r^2+a^2), 
\end{eqnarray}
where if we  express $n(r)$ in the form of  $n(r)\propto r^k$,  from which it follows that $k=0$ for $r\ll a$, and  $k=2$ for $r\gg a$.

\begin{figure*}[htbp]
	\gridline{\fig{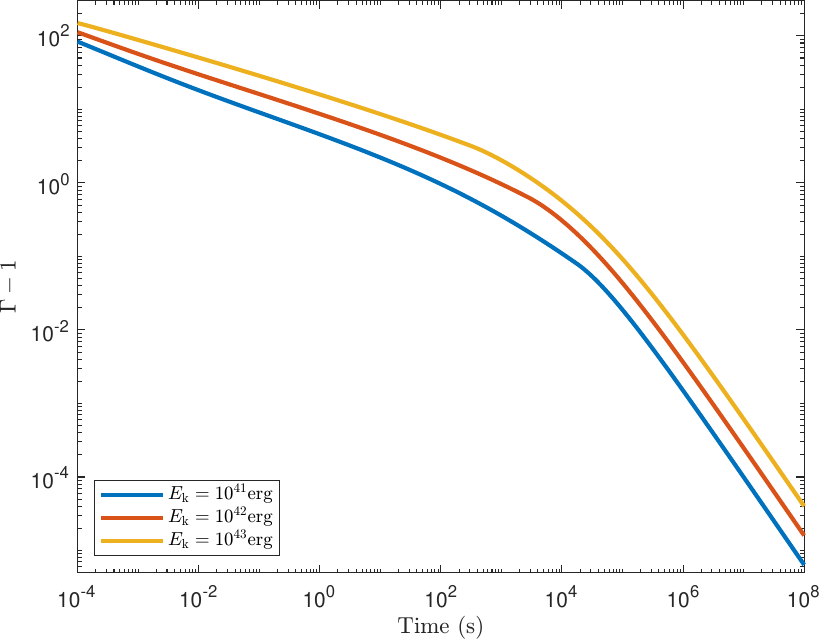}{0.5\textwidth}{(a)}
		\fig{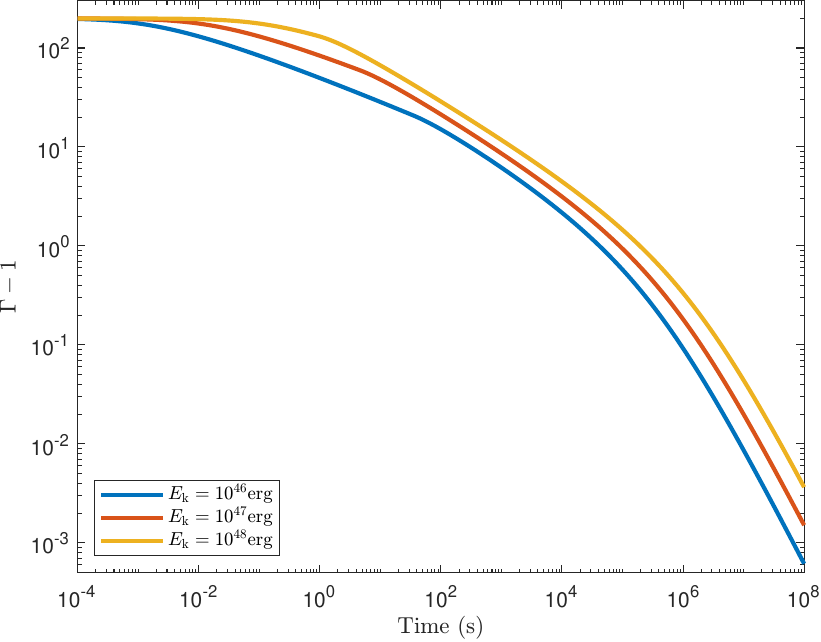}{0.5\textwidth}{(b)}}
	\caption{{Temporal evolution of the bulk Lorentz factor $\Gamma(t)-1$ for the two FRB cases. 
			Panel~(a): FRB~20200120E with $\dot{M}_{\rm w} = 10^{-11}~M_{\odot}~\mathrm{yr}^{-1}$ . 
			Panel~(b): FRB~20201124A with $\dot{M}_{\rm w} = 10^{-10}~M_{\odot}~\mathrm{yr}^{-1}$.}}
	\label{fig:7}
\end{figure*}

As the relativistic ejecta sweeps up material, it drives an external shock into the ambient medium, accelerating electrons and amplifying magnetic fields. 
The Lorentz factor $\Gamma$ of the shocked material at any given radius $r$ is determined by the energy conservation  \citep{10.1046/j.1365-8711.1999.02887.x}
\begin{eqnarray}
(\Gamma^2- 1) M_s c^2 +\frac{(\Gamma - 1)E_{\rm k}}{\eta-1}  = E_{\rm k}.
\end{eqnarray}
 The evolution of $\Gamma(r)$ depends on the density profile of the surrounding medi. {Here, we assume that the ejecta is cold prior to its interaction with the external medium. This assumption is commonly adopted in relativistic shock models for explosive transients and allows for a simplified treatment of the shock jump conditions. To clarify whether the shocked region remains relativistic throughout the time frame of interest or transitions to a Newtonian regime, we present in Figure \ref{fig:7} the evolution of the Lorentz factor $\Gamma$ as a function of time.}
Solve this quadratic equation in terms of $\Gamma$ and obtain its solutions as
\begin{eqnarray}
\Gamma = \frac{
	-E_{\mathrm{k}} + \sqrt{
		E_{\mathrm{k}}^2 + 4 M_{\mathrm{s}} c^2 E_{\mathrm{k}} \eta (\eta - 1) + 4 M_{\mathrm{s}}^2 c^4 (\eta - 1)^2
	}
}{
	2 M_{\mathrm{s}} c^2 (\eta - 1)~,
}
\end{eqnarray}
where 
{the initial Lorentz factor of the ejecta is taken as $\eta  = 200$, consistent with the FRB progenitor model adopted in this study \citep{Deng_2021,2021ApJ...917...13S}.  Such a high Lorentz factor is motivated by the need to allow coherent radio emission to escape the outflow without suppression by induced scattering, as required in several baryonic outflow models for FRBs. While the actual value of $\eta$ may vary depending on the specific progenitor system, our framework allows for flexibility. In Appendix, we also present afterglow calculations for a lower Lorentz factor ($\eta = 10$) to illustrate the impact of this parameter on the detectability of multi-wavelength afterglow emission.}
 
 \( M_s \) represents the mass inside a given radius \(r\), which is given by
\begin{eqnarray}
M_s(r) = \int_{0}^{r} n (r) m_p 4\pi r^2 \, dr.
\end{eqnarray}
Then, using the relation \citep{1997ApJ...489L..37S,Sari_1998}
\begin{eqnarray}
r = \frac{\beta ct}{1 - \beta},
\end{eqnarray}
one can derive the relationship between \(\Gamma\) and time \(t\) of the observer, where $\beta$ is the velocity of the shocked  material, $\Gamma=(1-\beta^2)^{-1/2}$.

\subsection{Synchrotron emission from forward shock}
As the relativistic ejecta expands, it interacts with the surrounding medium, driving a pair of shocks: a forward shock propagating into the ambient medium and a reverse shock traveling back into the ejecta. 
{However, in the parameter space considered in our model, the emission from  the reverse shock is not important (please see section 2.3).}
Given this short duration, we focus  on the afterglow emission produced by the forward shock.
The forward shock accelerates electrons and amplifies magnetic fields, producing synchrotron radiation observable at different wavelengths. The peak flux and spectral evolution of the afterglow depend on the magnetic field strength, electron energy distribution, and radiative cooling effects. We use the standard synchrotron radiation formalism following \citep{Sari_1998}.

We assume that the accelerated electron population follows a power-law distribution with an index {$p=2.2$} and that a fraction $\varepsilon_{\rm e}$ of the shock energy is transferred to electrons. The minimum Lorentz factor of injected electrons is given by
\begin{eqnarray}
	\gamma_{\rm m} = \varepsilon_{\rm e} \left( \frac{p - 2}{p - 1} \right) \frac{m_{\rm p}}{m_{\rm e}} (\Gamma - 1),
\end{eqnarray}
where \(m_{\rm p}\) and \(m_{\rm e}\) are the proton and electron masses, respectively. 
Moreover, by assuming that a fraction $\varepsilon_{\rm B}$ of the shock energy density is converted into magnetic energy, the post-shock magnetic field strength {in the fluid frame} is given by
\begin{eqnarray}
	B = \left( 8\pi e \varepsilon_{\rm B} \right)^{1/2},
\end{eqnarray} 
where $e$ represents the energy density in the shocked region. Drawing on the study on afterglows of gamma-ray bursts, $\epsilon_e = 0.1$ and $\epsilon_B = 10^{-4}$ are adopted in this  work  \citep{2015ApJS..219....9W}.

For synchrotron radiation, the power emitted by an electron with Lorentz factor $\gamma_e$ and its characteristic synchrotron frequency are given by
\begin{eqnarray}
	P(\gamma_e) &\simeq \frac{4}{3} \sigma_T c \Gamma^2 \gamma_e^2 \frac{B^2}{8\pi},
\end{eqnarray}
and
\begin{eqnarray}
	\nu(\gamma_e) &\simeq \Gamma \gamma_e^2 \frac{q_e B}{2\pi m_e c},
\end{eqnarray}
where \(q_e\) is the electron charge.

For an individual electron, the spectral power $P_\nu$ (in erg Hz$^{-1}$ s$^{-1}$) follows a broken power law, scaling as $\nu^{1/3}$ for $\nu < \nu(\gamma_e)$ and cutting off exponentially for $\nu > \nu(\gamma_e)$. The peak power occurs at $\nu(\gamma_e)$, with an approximate value 
\begin{eqnarray}
	P_{\nu,\max} \approx \frac{P(\gamma_e)}{\nu(\gamma_e)} = \frac{m_e c^2 \sigma_T}{3 q_e} \Gamma B.
\end{eqnarray}

\begin{figure*}[htbp]
	\gridline{\fig{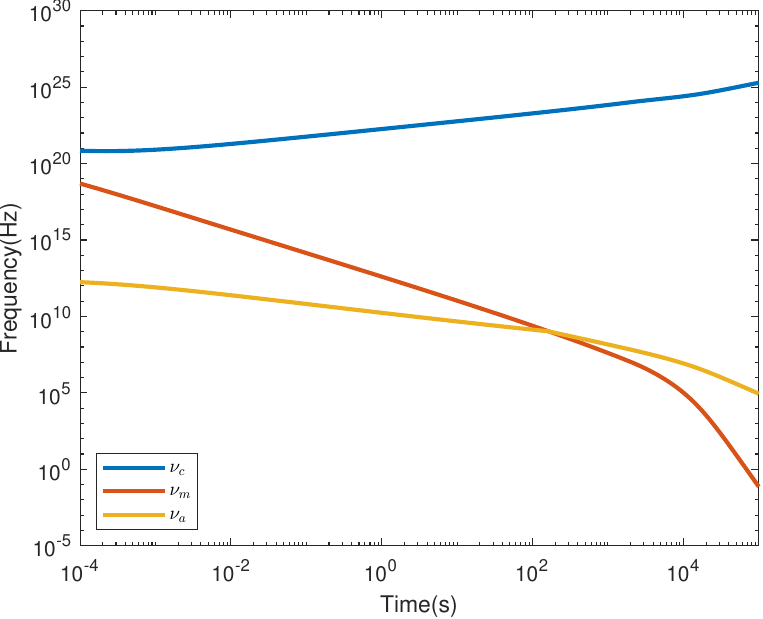}{0.5\textwidth}{(a)}
		\fig{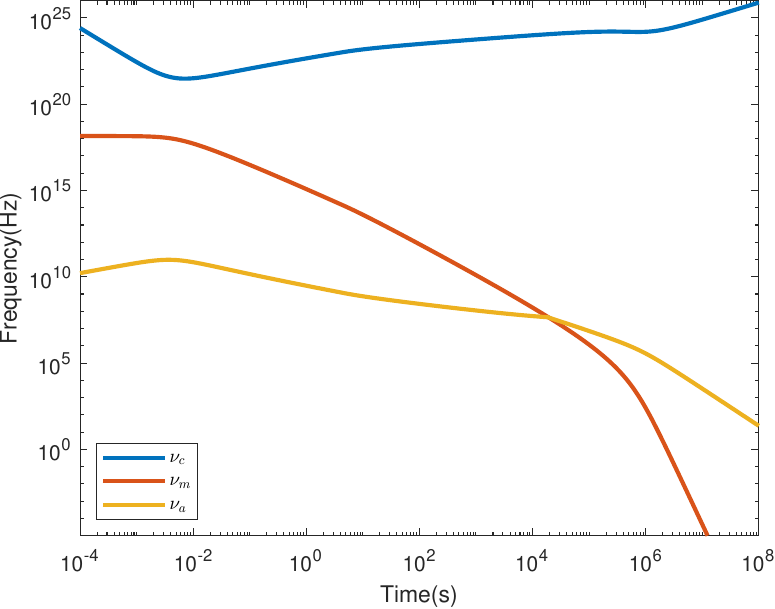}{0.5\textwidth}{(b)}}
	\caption{{Time evolution of the spectral break frequencies. 
			Panel~(a): FRB~20200120E with $\dot{M}_{\rm w} = 10^{-11}~M_{\odot}~\mathrm{yr}^{-1}$ and $E_{\rm k} = 10^{42}~\mathrm{erg}$. 
			Panel~(b): FRB~20201124A with $\dot{M}_{\rm w} = 10^{-10}~M_{\odot}~\mathrm{yr}^{-1}$ and $E_{\rm k} = 10^{47}~\mathrm{erg}$. 
			}}
	\label{fig:8}
\end{figure*}

The synchrotron radiation spectrum follows a multi-segment broken power law, characterized by key break frequencies: the minimum synchrotron frequency $\nu_m$, the cooling frequency $\nu_c$, and the self-absorption frequency $\nu_a$:

\begin{equation}
\nu_{\rm c} = \frac{\Gamma \gamma_c^2 q_e B}{2\pi m_e c},
\end{equation}

\begin{equation}
\nu_{\rm m} = \frac{\Gamma \gamma_m^2 q_e B}{2\pi m_e c},
\end{equation}

\begin{eqnarray}
    \nu_a = \left\{
    \begin{array}{ll}
        \left[ \frac{c_1 (p - 1)}{3 - k} \frac{q_e n(r) r}{B \gamma_m^5} \right]^{3/5} \nu_m, & \text{if } \nu_a < \nu_m, \\
        \left[ \frac{c_2 (p - 1)}{3 - k} \frac{q_e n(r) r}{B \gamma_m^5} \right]^{2/(p + 4)} \nu_m, & \text{if } \nu_m < \nu_a < \nu_c,
    \end{array}
    \right.
\end{eqnarray}
where $c_1, c_2$ are coefficients that depend on $p$ \citep{2018pgrb.book.....Z}. $k=0$ for $r\ll a$, and $k=2$ for $r\gg a$, according to Eq.(3).  {It can be seen from Figure \ref{fig:8} that the process has been in a slow cooling state  over the relevant timescales. }

The peak specific luminosity measured by the observer can be  estimated as
\begin{eqnarray}
	L_{\rm \nu,max} \sim N_e P_{\nu,\max},
\end{eqnarray}
where \(N_e\) is  the total number of radiating electrons in the downstream of shock.
Then, the  specific luminosity measured by the observer  is given by \citep{Sari_1998}
\begin{equation}
	\begin{aligned}
		L_{\nu} &= \left\{
		\begin{array}{ll}
			(\nu/\nu_a)^2(\nu_a/\nu_m)^{1/3} L_{\nu,\max}, & \nu < \nu_a, \\
			(\nu/\nu_m)^{1/3} L_{\nu,\max}, & \nu_a < \nu < \nu_m, \\
			(\nu/\nu_m)^{-(p-1)/2} L_{\nu,\max}, & \nu_m < \nu < \nu_c, \\
			(\nu/\nu_c)^{-p/2} (\nu_c/\nu_m)^{-(p-1)/2} L_{\nu,\max}, & \nu_c < \nu ,
		\end{array}
		\right.
	\end{aligned}
\end{equation}
for \(\nu_a < \nu_m < \nu_c\), and
\begin{equation}
	L_{\nu} = \left\{
	\begin{array}{ll}
		(\nu/\nu_m)^2 (\nu_m/\nu_a)^{(p+4)/2} L_{\nu,\max}, & \nu < \nu_m, \\
		(\nu/\nu_a)^{5/2} (\nu_a/\nu_m)^{-(p-1)/2} L_{\nu,\max}, & \nu_m < \nu < \nu_a, \\
		(\nu/\nu_m)^{-(p-1)/2} L_{\nu,\max}, & \nu_a < \nu < \nu_c, \\
		(\nu/\nu_c)^{-p/2} (\nu_c/\nu_m)^{-(p-1)/2} L_{\nu,\max}, & \nu_c < \nu,
	\end{array}
	\right.
\end{equation}
for \(\nu_m < \nu_a < \nu_c\).
This spectral framework provides a robust description of synchrotron emission from the forward shock, capturing its temporal and spectral evolution under various radiative regimes.

\subsection{The emission from reverse shock}
{In this section, we examine the properties of the reverse shock with the aim of clarifying why its emission is expected to be of limited significance within the framework of this study. In the stellar wind environment considered here, substantial reverse shock radiation can arise only if the ambient medium is dense enough for the shock to become relativistic \citep{2003MNRAS.342.1131W}. In the absence of such conditions, its contribution is negligible compared to that of the forward shock. Therefore, our discussion below focuses on the specific case in which the reverse shock is relativistic.
}

{For a relativistic reverse shock to develop, the deceleration timescale of the ejecta, $t_{\rm dec}$, must be shorter than the prompt emission duration of the FRB, $\delta t \sim 10^{-3}$ s. In our analysis, this criterion is met for FRB 20200120E, and for FRB 20201124A in cases where the ambient density is sufficiently high. As noted above, only under such circumstances can reverse shock emission be appreciable. Nevertheless, we will show below that even in the relativistic regime, the resulting afterglow emission is negligible. In this situation, the reverse shock crossing radius is
$r_{\times} < 10^{11} \eta_2^2 \delta t_{-3}~\mathrm{cm}$,
which is smaller than the orbital separation $a$ adopted in this work. Consequently, during the entire reverse shock crossing, the density of the surrounding medium remains approximately constant, $n(r) \simeq \mathrm{const.}$, as given by Eq. (3).
At the reverse shock crossing time $t_\times = \delta t$, the ratio of the peak specific flux from the reverse shock to that from the forward shock is \citep{2018pgrb.book.....Z}
\begin{equation}
\frac{F^{\rm r}_{\nu,\max}}{F^{\rm f}_{\nu,\max}} \simeq \Gamma_{\times} \mathcal{R}_{\rm B},
\end{equation}
where $\Gamma_{\times}$ is the bulk Lorentz factor at $t = t_\times$, and $\mathcal{R}_{\rm B}^2$ is the ratio of magnetic equipartition factors $\epsilon_{\rm B}$ between the reverse and forward shocks. Since the ejecta is considered to be baryonic in this study,  one can expect  $\mathcal{R}_{\rm B} \sim 1$. Thus, we have $ F^{\rm r}_{\rm \nu,max}/F^{\rm f}_{\rm \nu,max}\lesssim 10^{2}$.
}

{For $t > t_\times$, this flux ratio declines as $t^{-9/8}$ for a wind-like medium ($k=2$; \citealt{2003MNRAS.342.1131W,2018pgrb.book.....Z}) and as $t^{-47/48}$ for a uniform medium ($k=0$; \citealt{Kobayashi_2000,2018pgrb.book.....Z}). Consequently, by $t \gtrsim 0.1~\mathrm{s}$, the reverse shock peak flux already falls below that of the forward shock, i.e., $F^{\rm r}_{\nu,\max} \lesssim F^{\rm f}_{\nu,\max}$.
In other words, on observational timescales longer than subsecond, the reverse shock emission has already decayed to a level significantly weaker than that of the forward shock. As a result, the reverse shock emission is either characterized by an extremely short observable duration or is intrinsically much fainter than the forward shock emission. For this reason, we exclude the reverse shock contribution from further consideration in our modeling. This conclusion, as evident from the above analysis, arises from the intrinsically short timescale of FRB prompt emission ($\delta t \sim 10^{-3}~\mathrm{s}$). Therefore, in most FRB afterglow scenarios, the reverse shock emission is expected to be negligible.
}

\section{Case Study} \label{Sec.3}
To explore the observational signatures of FRB afterglows, we conduct case studies on two well-characterized repeating fast radio bursts: FRB 20200120E and FRB 20201124A. For each case, we model the potential multi-wavelength afterglow light curves, assess their detectability, and propose optimal follow-up strategies. {In our study, we focus on the brightest bursts from active repeating FRBs, which are the most likely to produce detectable afterglows. While repeated bursts may increase the total energy released over time, our analysis is centered on the brightest bursts, which typically have much longer repetition timescales of their afterglows. {To facilitate reference to the basic observational properties of the two FRBs analyzed in this study, we summarize their key parameters—including distance, fluence of the brightest detected bursts, and relevant recurrence-rate estimates—in Table~\ref{tab:FRB_params}.}}

{While the study focuses on the non-thermal X-ray afterglow resulting from the interaction of relativistic ejecta with the surrounding medium, we acknowledge that thermal X-ray emission from the accretion disk or hot spots could also contribute to the X-ray emission. However, given the distinct temporal evolution of the two components, the thermal X-ray emission is expected to be separate from the afterglow contribution. This thermal emission could potentially overlap with the X-ray afterglow, but it is beyond the scope of our current study, which primarily models the afterglow from the relativistic ejecta. Future work could explore the interplay between these two emission sources in more detail.}
    
\subsection{FRB 20200120E}
FRB 20200120E is a repeating FRB located in a globular cluster of the nearby galaxy M81 \citep{2022Natur.602..585K}, at a distance of 3.6 Mpc. Its proximity makes it an excellent candidate for afterglow searches.
Given that its host environment is a globular cluster, the ambient density is expected to be low. We adopt a typical interstellar medium density of  $n_{\rm I} \sim 10^{-2} \rm cm^{-3}$ within globular clusters \citep{2001ApJ...557L.105F}. 
The binary parameters, including wind velocity $v_{\rm w} = 10^{8}~ \rm cm~s^{-1}$, orbital period $T = 0.1$ days, and companion star mass $M_{\rm C} = 0.1 M_{\odot}$, are chosen to represent a typical low-mass binary containing a neutron star \citep{2022abn..book.....C}\footnote{It should be noted that appropriately adjusting the specific values of these parameters within the parameter range for low-mass binaries will not alter the conclusions of this paper. The conclusions remain robust within a reasonable range of parameter variations.}.
The choice of kinetic energy values is motivated by observational constraints. The most energetic radio burst detected from FRB 20200120E has an isotropic-equivalent energy of approximately $10^{36}$ erg \citep{2024NatCo..15.7454Z}. Assuming a radiation efficiency similar to that of the Galactic FRB 20200428D, which is about $10^{-5}$ \citep{2020ApJ...899L..27M}, we estimate that the ejecta kinetic energy could be as high as $10^{41}$ erg. Therefore, we adopt $E_{\rm k} = 10^{41}$, $10^{42}$, and $10^{43}$ erg to explore a reasonable range of afterglow possibilities.

Figure \ref{fig:1} presents the afterglow light curves for FRB 20200120E. In this case, a typical mass-loss rate, for a compact pulsar binary, $\dot{M}_{\rm w}=10^{-11} M_{\odot} ~\rm yr^{-1}$ is adopted \citep{2022abn..book.....C,2023Natur.620..961P}. Panel (a) shows the radio afterglow at 1 GHz. The peak flux densities for different kinetic energy values are \(5.47 \times 10^{-8}\) Jy (\(E_{\rm k} = 10^{41}\) erg, green), \(2.26 \times 10^{-7}\) Jy (\(E_{\rm k}= 10^{42}\) erg, blue), and \(6.02 \times 10^{-7}\) Jy (\(E_{\rm k} = 10^{43}\) erg, red). 
{The dashed purple line marks the detection sensitivity of the Square Kilometre Array (SKA), evaluated using the standard radiometer equation \citep{2019arXiv191212699B}.  
The sensitivity is scaled with observing time as $t_{\rm obs}^{-1/2}$, assuming that the integration time increases with the observation timescale. 
}
 It is evident that future SKA observations could potentially detect the radio afterglow associated with bright bursts of FRB 20200120E.
Panel (b) displays the optical afterglow in the R band. The peak magnitudes are 22.0 (\(E_{\rm k} = 10^{41}\) erg, green), 21.0 (\(E_{\rm k} = 10^{42}\) erg, blue), and 20.2 (\(E_{\rm k} = 10^{43}\) erg, red). The detection sensitivity of Large Synoptic Survey Telescope (LSST) is indicated by the dashed purple line {in the survey mode, which reaches 24.5mag in 30 s\citep{2014ApJ...792L..21Y}}. 
It can be seen that due to the insufficient peak brightness of the optical afterglow and the extremely short duration at peak intensity, the optical afterglow is not practically detectable.
Panel (c) shows the X-ray afterglow at 1 keV.  Even in the most optimistic case ($E_{\rm k} = 10^{43}$ erg), the afterglow remains below the detection sensitivity of Swift/XRT {which is $\propto t^{-1}$ early on and breaks to $\propto t^{-1 / 2}$ , where $ F_{v}=2.0 \times 10^{-15} \mathrm{erg} \mathrm{~cm}^{-2} \mathrm{~s}^{-1}$ at $t = 10^5s$\citep{Moretti2008MCAONP}}.

To explore the impact of the surrounding environment on the afterglow brightness, Figure \ref{fig:2} presents the light curves for different stellar wind mass-loss rates of the companion star ($\dot{M}_{\rm w} = 10^{-10}$, $10^{-9}$, and $10^{-8}~M_{\odot}\rm yr^{-1}$), with a  fiducial ejecta kinetic energy of $E_{\rm k}= 10^{41}$ erg, while keeping all other parameters the same as in Figure \ref{fig:1}.
Panel (a) shows the radio afterglow at 1 GHz. The peak flux densities for different mass-loss rates are $1.65 \times 10^{-7}$ Jy ($\dot{M}_{\rm w} = 10^{-10}M_{\odot}\rm yr^{-1}$, green), $6.17 \times 10^{-7}$ Jy ($\dot{M}_{\rm w} = 10^{-9}M_{\odot}\rm yr^{-1}$, blue), and $2.17 \times 10^{-6}$ Jy ($\dot{M}_{\rm w} = 10^{-8}M_{\odot}\rm yr^{-1}$, red).  Higher mass-loss rates result in brighter radio afterglows, but even in the most optimistic scenario, detection is only feasible with SKA-class instruments.
Panel (b) presents the optical afterglow in the R band. The peak magnitudes are 19.6 ($\dot{M}_{\rm w} = 10^{-10}M_{\odot}\rm yr^{-1}$, green), 17.9 ($\dot{M}_{\rm w} = 10^{-9}M_{\odot}\rm yr^{-1}$, blue), and 16.4 ($\dot{M}_{\rm w} = 10^{-8}M_{\odot}\rm yr^{-1}$, red). It can be seen that the optical afterglow is highly sensitive to the environmental density. That is, a higher rate of mass loss significantly enhances the brightness of the optical afterglow.
These  peak magnitudes are below the LSST sensitivity limit, suggesting that optical afterglows may be detectable. However, the duration at peak intensity is extremely short, and the brightness decays very rapidly. Consequently, even a sophisticated facility like LSST is unlikely to conduct effective observations. Only Bright bursts occurring in high-density environments can produce optical afterglows that reach detectable levels. Anyhow effective optical observations can always serve as a key probe for constraining the characteristics of the surrounding environment. 
Panel (c) illustrates the X-ray afterglow at 1 keV. Even under the most optimistic conditions, the X-ray afterglow remains below the Swift/XRT detection limit.

To facilitate direct comparison with radio observations, we include light curves at 5 GHz and 500 MHz, as shown in Figure \ref{fig:5}. Panels (a), (b), and (c) correspond to three different energy parameters, while panels (d), (e), and (f) represent cases with varying stellar wind parameters $\dot{M}_{\rm w}$. {To provide a compact overview of the detectability of the afterglow across the relevant parameter space, we construct two-dimensional contour maps in the $(E_{\rm k}, \dot{M}_{\rm w})$ plane. In figure \ref{fig:9}, the color scale represents the maximum spectral flux density reached over the entire light curve. The solid line indicates the instrumental sensitivity threshold, above which the afterglow becomes detectable. Dashed lines mark the latest observation times at which the flux remains above the sensitivity limit with different colors.This representation allows the peak brightness, the detectable parameter region,and the characteristic observational time window to be assessed simultaneously.}

Based on the predicted afterglow light curves of FRB 20200120E, we propose an observational strategy to optimize detection across multiple wavelengths. Radio observations with the future SKA offer the best chance of detecting the afterglow, as the predicted flux densities, particularly for bright bursts within high-density environments, fall within SKA's sensitivity range. 
In the optical band, although the predicted afterglow is typically too faint and short-lived for detection by facilities like LSST, bright bursts occurring in dense environments can significantly enhance the brightness to detectable levels.
Given the rapid decay of optical afterglows, continuous monitoring or high-cadence surveys are required, with targeted follow-up immediately after bright FRB detections offering the best chance of success. The X-ray afterglow detection is unlikely with current instruments such as Swift/XRT. Overall, SKA observations should be prioritized, while optical monitoring remains a promising secondary approach under specific conditions, and X-ray follow-up should be considered for future instrumentation with improved sensitivity.

\subsection{FRB 20201124A}
FRB 20201124A is a highly active and energetic repeating FRB. The strongest burst from this source had a fluence of 640 Jy ms and an isotropic energy release of  \(10^{41}\) erg \citep{2022MNRAS.512.3400K}.
Given the high energetics of FRB 20201124A, the potential afterglow associated with the bright burst may still be detectable at a distance of 453 Mpc in future observations, making this FRB an ideal candidate for afterglow searches. 
Therefore, following the approach described for FRB 20200120E, we performed multi-wavelength afterglow calculations for FRB 20201124A.  
Similarly, based on the assumption of a radiation efficiency of \(10^{-5}\), we consider the following possible kinetic energies for the ejecta: \(E_{\rm k}  = 10^{46}\), \(10^{47}\), and \(10^{48}\) erg. 
The binary parameters, including wind velocity $v_{\rm w} = 10^{8} \rm~ cm~s^{-1}$, orbital period $T = 80$ days, and companion star mass $M_{\rm C} = 8.0M_{\odot}$, are adopted from  \cite{2022NatCo..13.4382W}.  In this scenario, given that we are here considering  that FRB 20201124A is in a massive binary system, thus we set $n_{\rm I}=1$, which represents the typical interstellar medium density of a normal galaxy.

Figure \ref{fig:3} presents the multi-wavelength afterglow light curves of FRB 20201124A for three different ejecta kinetic energy values: \(E_{\rm k} = 10^{46} \, \rm{erg}\) (green, \(E_{\rm k}= 10^{47} \, \rm{erg}\) (blue), and \(E_{\rm k} = 10^{48} \, \rm{erg}\) (red).  In this case, a referenced mass loss rate for the companion star  $\dot{M}_{\rm w}=10^{-10} M_{\odot} \rm yr^{-1}$ is adopted \citep{2022NatCo..13.4382W}.
 Panel (a) shows the radio afterglow at 1 GHz. The peak flux densities for different kinetic energies are \(2.74 \times 10^{-8} \, {\rm Jy}\) (\(E_{\rm k}  = 10^{46} \, \rm{erg}\)), \(1.84 \times 10^{-7} \, {\rm Jy}\) (\(E_{\rm k}  = 10^{47} \, \rm{erg}\)), and \(1.54 \times 10^{-6} \, {\rm Jy}\) (\(E_{\rm k}  = 10^{48} \, \rm{erg}\)). 
Panel (b) presents the optical afterglow in the R band.  The peak magnitudes for different kinetic energy values are 25.5 (\(E_{\rm k}  = 10^{46} \, \rm{erg}\)), 24.7 (\(E_{\rm k}  = 10^{47} \, \rm{erg}\)), and 22.6 (\(E_{\rm k}  = 10^{48} \, \rm{erg}\)).  Panel (c) displays the X-ray afterglow at 1 keV.  
Among these, the radio afterglow is most sensitive to the kinetic energy of the ejecta, indicating that radio follow-up observations provide the best opportunity to detect the afterglow associated with FRB 20201124A. 
Optical afterglows could be detectable by LSST in the case of extremely bright bursts, though the short radiation timescale presents a challenge. In contrast, X-ray afterglows are unlikely to be observed with current instruments.

Figure \ref{fig:4} presents the light curves of the multi-wavelength afterglow of FRB 20201124A under varying mass loss rates of stellar winds from the companion star, with a  fiducial ejecta kinetic energy of $E_{\rm k} = 10^{46}$ erg.  
Three different wind mass-loss rates  are considered: \(\dot{M}_{\rm w} = 10^{-10} M_{\odot}\rm yr^{-1}\) (green),  \(\dot{M}_{\rm w} = 10^{-8} M_{\odot}\rm yr^{-1}\) (blue), and  \(\dot{M}_{\rm w} = 10^{-6} M_{\odot}\rm yr^{-1}\) (red).  The other corresponding model parameters are the same as those in Figure \ref{fig:3}.
Panel (a) shows the   light curves of the radio afterglows in 1 GHz, where  the peak flux densities for different mass-loss rates are \(2.74 \times 10^{-8} \, {\rm Jy}\) (\(\dot{M}_{\rm w} = 10^{-10} M_{\odot}\rm yr^{-1}\)), \(1.33 \times 10^{-7} \, {\rm Jy}\) (\(\dot{M}_{\rm w} = 10^{-8} M_{\odot}\rm yr^{-1}\)), and \(8.03 \times 10^{-7} \, {\rm Jy}\) (\(\dot{M}_{\rm w} = 10^{-6} M_{\odot}\rm yr^{-1}\)). 
Panel (b) presents the optical afterglow in R band. The peak magnitudes for \(\dot{M}_{\rm w} = 10^{-10}M_{\odot}\rm yr^{-1}\), \(\dot{M}_{\rm w} = 10^{-8}M_{\odot}\rm yr^{-1}\), and \(\dot{M}_{\rm w} = 10^{-6}M_{\odot}\rm yr^{-1}\) are 25.4, 20.3, and 15.6, respectively. Panel (c) shows the X-ray afterglow in 1 keV band.
These results indicate that a denser circumstellar environment, created by a high stellar wind mass-loss rate, significantly enhances the optical afterglow brightness. If the companion star has a strong wind, the optical afterglow could be  much brighter than the detection limit of LSST.  Therefore, optical monitoring or follow-up observations are also worthwhile.  And the optical observations may serve as an effective probe for constraining the density of the surrounding environment FRB 20201124A.
The X-ray afterglow remains faint even for the highest mass-loss rate, reinforcing the conclusion that X-ray observations are unlikely to be successful with current telescopes.  

Similar to Figure \ref{fig:5}, we also include 5 GHz and 500 MHz radio light curves for FRB 20201124A to facilitate direct comparison with radio observations, as shown in Figure \ref{fig:6}. Panels (a), (b), and (c) correspond to three different energy parameters, while panels (d), (e), and (f) represent cases with varying stellar wind parameters $\dot{M}_{\rm w}$.  
{Following the same approach, we summarize the detectability of the afterglow from FRB~20201124A using contour maps in the $(E_{\rm k}, \dot{M}_{\rm w})$ parameter space, shown by figure \ref{fig:10}. As in the case of FRB~20200120E, the color contour shows the flux density achieved over the full light curve across the parameter space, while the solid line denotes the instrumental sensitivity threshold. Dashed lines indicate the latest time at which the afterglow remains detectable. These maps provide a direct comparison between different physical regimes and clearly illustrate how the observable time window depends on the ejecta energy and the stellar wind environment.}

{
Recent observations have also suggested the presence of a persistent radio source associated with FRB~20201124A \citep{2024Natur.632.1014B}, which may complicate the detectability of a transient afterglow component. The reported flux density of this source is $\sim 8.2,\mu$Jy at 6~GHz, with an inverted spectrum characterized by a spectral index of $\alpha \simeq 1$. Extrapolating to lower frequencies yields flux densities of $\sim 1.4,\mu$Jy at 1~GHz and $\sim 0.3,\mu$Jy at 500~MHz.
 To account for this effect, in Figure~10 we explicitly indicate the parameter-space regions where the predicted afterglow peak flux exceeds the persistent emission level. The red solid curve marks the threshold above which the 1~GHz afterglow is expected to outshine the persistent source, while the blue solid curve shows the corresponding threshold at 500~MHz. As illustrated, at 1~GHz only a limited region of the $(E_{\rm k}, \dot{M}_{\rm w})$ parameter space satisfies this condition, whereas at 500~MHz a substantially larger fraction remains detectable.
 This comparison demonstrates that, although persistent radio emission can obscure the afterglow in part of the parameter space, especially at gigahertz frequencies, lower-frequency observations provide a more favorable window for isolating the transient afterglow signal from FRB~20201124A.
}

Based on our analysis above, we propose the following observational strategies for detecting the afterglow of FRB 20201124A.
Since the radio afterglow is highly sensitive to the ejecta kinetic energy, long-term radio monitoring provides the most promising opportunity for detection, particularly in the parameter space with relatively high kinetic energy and/or dense circumstellar environments. A higher stellar wind mass-loss rate ($\dot{M}{\rm w} = 10^{-6} M{\odot}\mathrm{yr}^{-1}$) can significantly enhance the radio flux, making high-sensitivity facilities such as the future SKA or existing instruments like MeerKAT well suited for follow-up observations. Optical afterglow detection is feasible for bright bursts, especially with LSST-class telescopes. For sufficiently high ejecta kinetic energy, the peak $R$-band magnitude can approach the LSST detection limit, and a dense stellar wind environment can further increase the optical brightness. Therefore, rapid optical follow-up observations are strongly encouraged, as even nondetections can place useful constraints on the properties of the surrounding medium. In contrast, X-ray afterglows remain faint across the explored parameter space and are unlikely to be detectable with current instruments such as Swift/XRT, even under optimistic conditions. Consequently, X-ray follow-up is not expected to be effective for FRB~20201124A.

\section{SUMMARY}  \label{Sec.5}	
In this work, we have developed a unified framework to predict the multi-wavelength afterglows associated with FRBs originating in binary systems. Our analysis focuses on two representative sources, FRB 20200120E and FRB 20201124A, examining how their ejecta interact with the circumstellar medium shaped by the companion star’s stellar wind. By computing the expected light curves in the radio, optical, and X-ray bands, we assess the detectability of these afterglows with current and future observational facilities. 

Our results indicate that radio afterglows are the most promising observational signature of FRB afterglows. The radio flux is highly sensitive to both the kinetic energy of the ejecta and the density of the circumstellar medium. Higher ejecta energies and denser environments lead to stronger radio emission, making these afterglows detectable with high-sensitivity radio telescopes such as the SKA and MeerKAT. Future deep radio observations will be crucial for confirming the existence of FRB afterglows and constraining the nature of the FRB progenitor environment. Additionally, detecting radio afterglows will provide valuable insights into the energetics and dynamics of FRB ejecta, which are key to understanding their physical origin.

Optical afterglows, while potentially detectable under favorable conditions, face significant observational challenges. Our calculations show that a high mass-loss rate from the companion star can significantly enhance optical emission. However, optical afterglows are typically short-lived and decay rapidly, making them difficult to capture with traditional survey strategies. High-cadence optical survey or  monitoring, such as with the LSST, could play a crucial role in probing the FRB environment. Even non-detections in the optical band would help place constraints on the density of the circumstellar medium.

A dense surrounding medium can enhance afterglow brightness, increasing the likelihood of detecting optical emission. If FRB 20200120E resides in a spider pulsar system—common in globular clusters—its companion star, being in close proximity to the neutron star, may exhibit strong stellar winds. In such a scenario, monitoring the optical afterglow of FRB 20200120E would be highly intriguing. Likewise, if FRB 20201124A originates in a massive binary system with a Be star companion, the rapid rotation of the Be star could result in an exceptionally high mass-loss rate during its active phase. Our results suggest that, in this case, the optical afterglow of FRB 20201124A would be significantly brighter and longer-lasting than that of FRB 20200120E, making it an excellent target for optical afterglow searches. Dedicated optical monitoring of FRB 20201124A thus presents a valuable opportunity for afterglow detection.  

In contrast, the X-ray afterglows of both FRB 20200120E and FRB 20201124A are predicted to be too faint to be detected with current X-ray instruments such as Swift/XRT. This suggests that only next-generation high-sensitivity X-ray telescopes will be capable of placing meaningful constraints on FRB afterglows in the X-ray regime.

Overall, future radio and optical follow-up campaigns will be essential for detecting afterglows and advancing our understanding of FRBs. If FRBs indeed originate from binary systems, as suggested for FRB 20200120E and FRB 20201124A, this implies that the surrounding medium plays a crucial role in shaping their afterglow properties. Multi-wavelength observations of FRB afterglows will therefore serve as a powerful diagnostic tool for probing FRB progenitors and their environments.

\section*{Acknowledgments}
This work is supported by the National Natural Science Foundation of China (grant Nos. 12203013 and 12494575), and the Guangxi Talent Program (Highland of Innovation Talents).
\FloatBarrier
\bibliography{References}
\bibliographystyle{aasjournal}

\begin{table}[htbp]
	\centering
	\caption{{Basic observational properties of the FRBs analyzed in this study.}}
	\begin{tabular}{lcccccc}
		\hline\hline
		Source  &  Distance (Mpc) &Time of arrival (MJD)& Fluence (Jy ms) &Energy (erg)& Repetition &Reference \\
		\hline
		FRB 20200120E & 3.6 &59473.6107315& 30 & $\sim  10^{36}$ & $>0.5 ~$yr$^{-1}$ &  \cite{2024NatCo..15.7454Z}\\
		FRB 20201124A &453 &59305.47976873 & 640&$\sim 10^{41}$ &  $\sim 0.3~$yr$^{-1}$&  \cite{2022MNRAS.512.3400K}  \\
		\hline
	\end{tabular}
\tablecomments{{The repetition listed here refer to the occurrence rates of bursts with fluences comparable to the brightest events considered in this work.
		For FRB~20201124A, the rate is inferred from the cumulative energy distribution reported by \cite{2022RAA....22l4002Z}, which follows $N(>E)\propto E^{-3.27}$ above $10^{39} \mathrm{erg}$.
		For FRB~20200120E, we estimate a conservative lower limit based on the single bright-burst detection during the 62.5-hour monitoring campaign of \cite{2024NatCo..15.7454Z}, incorporating Poisson uncertainties \cite{1986ApJ...303..336G}.
}}
	\label{tab:FRB_params}
\end{table}

\begin{figure*}[htbp]
	\gridline{\fig{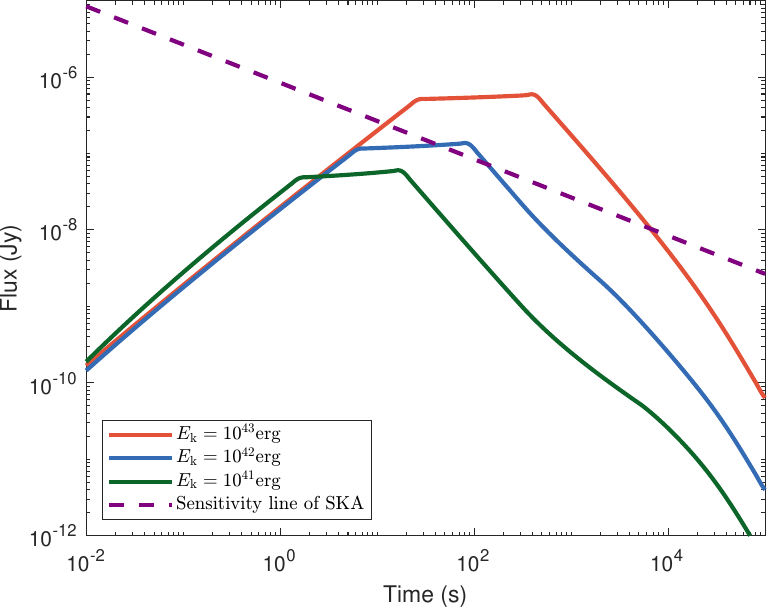}{0.3\textwidth}{(a)}
		\fig{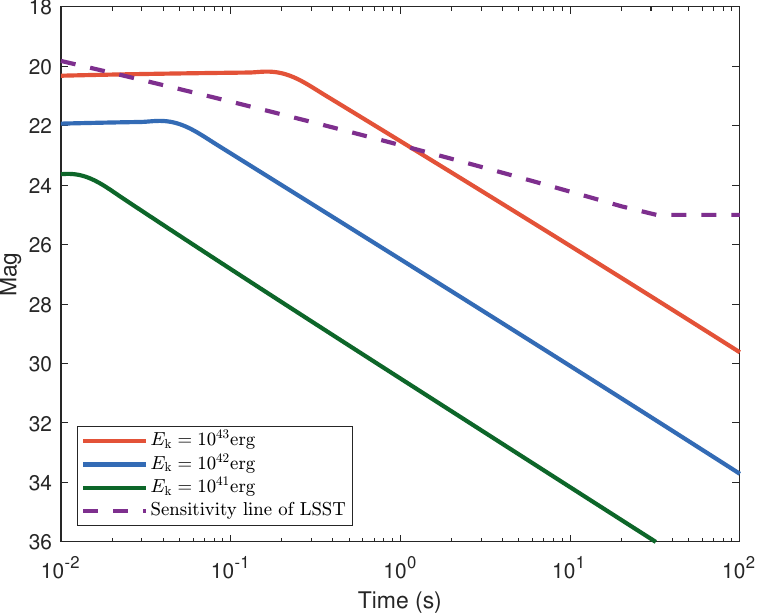}{0.29\textwidth}{(b)}
		\fig{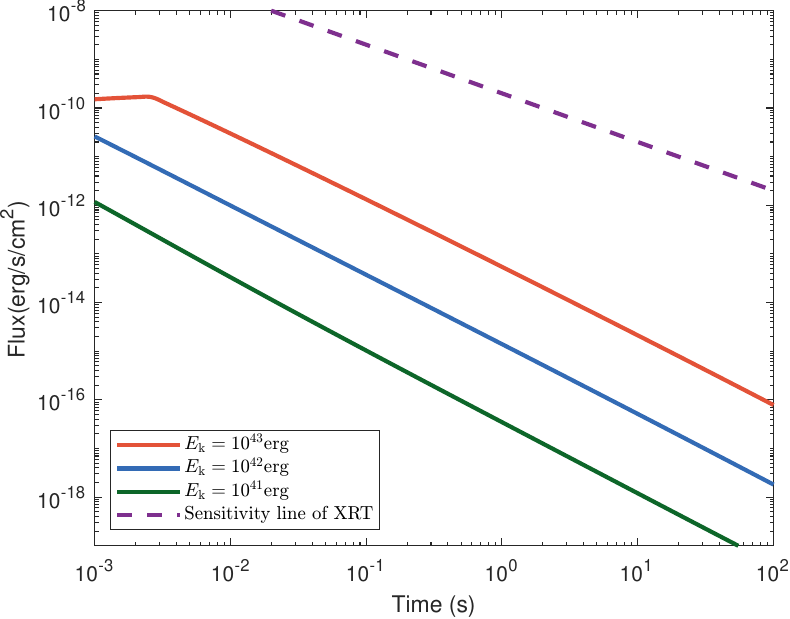}{0.3\textwidth}{(c)}}
	\caption{{Multi-wavelength afterglow light curves of FRB~20200120E for different ejecta kinetic energies: 
			$E_{\rm k} = 10^{41}~\mathrm{erg}$ (green), $10^{42}~\mathrm{erg}$ (blue), and $10^{43}~\mathrm{erg}$ (red). The stellar-wind mass-loss rate is fixed at $\dot{M}_{\rm w} = 10^{-11} M_{\odot} \mathrm{yr}^{-1}$ in all three panels.
			Panel~(a): 1~GHz radio afterglow with SKA sensitivity limit (dotted purple; \citealt{2019arXiv191212699B}). 
			Panel~(b): $R$-band optical afterglow with LSST sensitivity limit (dotted purple; \citealt{2014ApJ...792L..21Y}). 
			Panel~(c): 1~keV X-ray afterglow with Swift/XRT sensitivity limit (dotted purple; \citealt{Moretti2008MCAONP}).}}
	\label{fig:1}
\end{figure*}

\begin{figure*}[htbp]
	\gridline{\fig{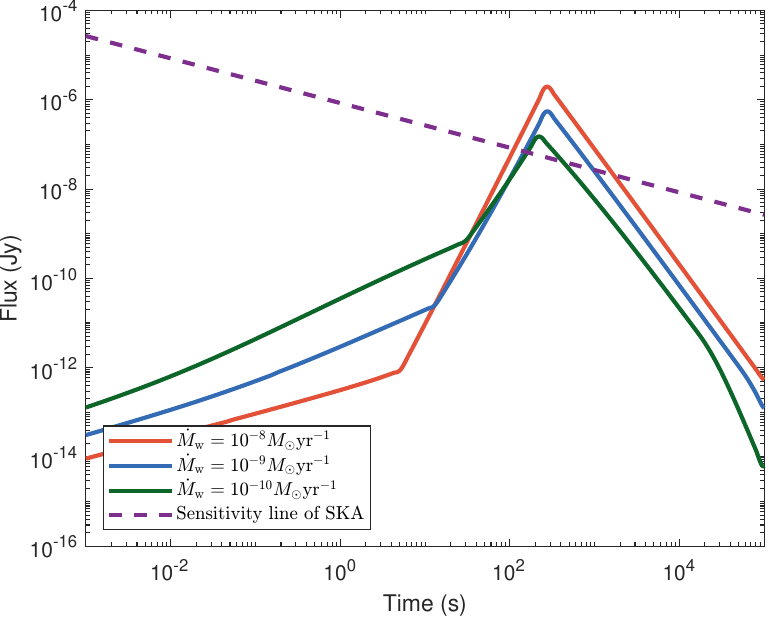}{0.3\textwidth}{(a)}
		\fig{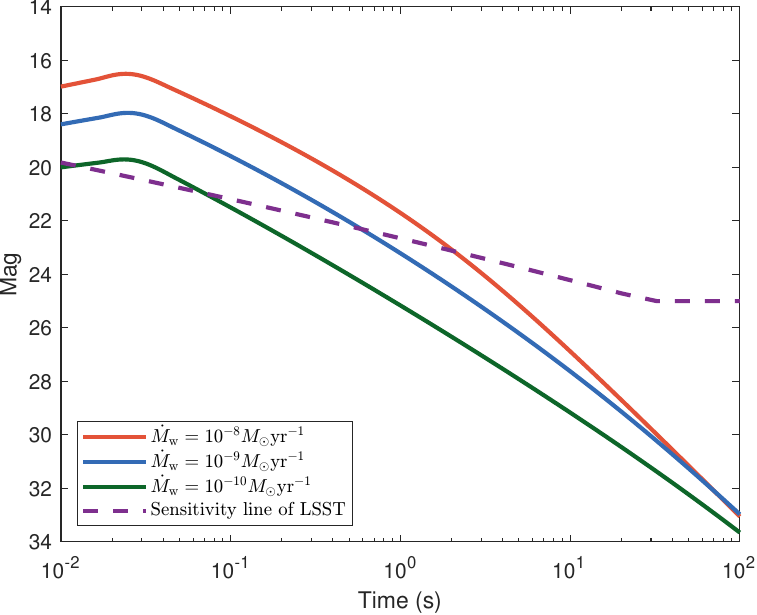}{0.29\textwidth}{(b)}
		\fig{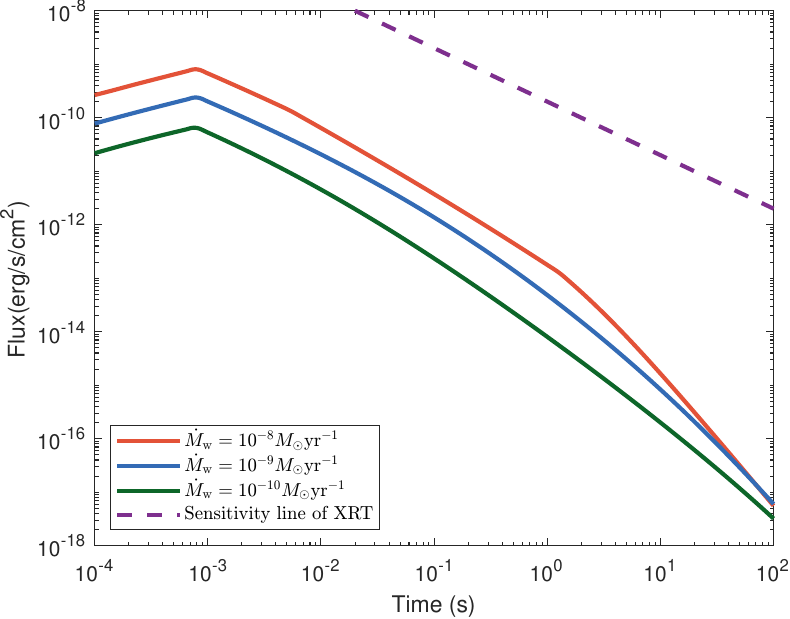}{0.3\textwidth}{(c)}}
	\caption{{Multi-wavelength afterglow light curves of FRB~20200120E for different stellar wind mass-loss rates: 
			$\dot{M}_{\rm w} = 10^{-10}~M_{\odot}~\mathrm{yr}^{-1}$ (green), $10^{-9}~M_{\odot}~\mathrm{yr}^{-1}$ (blue), and $10^{-8}~M_{\odot}~\mathrm{yr}^{-1}$ (red). 
			Panels as in Figure~\ref{fig:1}. The ejecta kinetic energy is fixed at $E_{\rm k} = 10^{41} \mathrm{erg}$ in all three panels.}}
	\label{fig:2}
\end{figure*}

\begin{figure*}[ht]
	\gridline{\fig{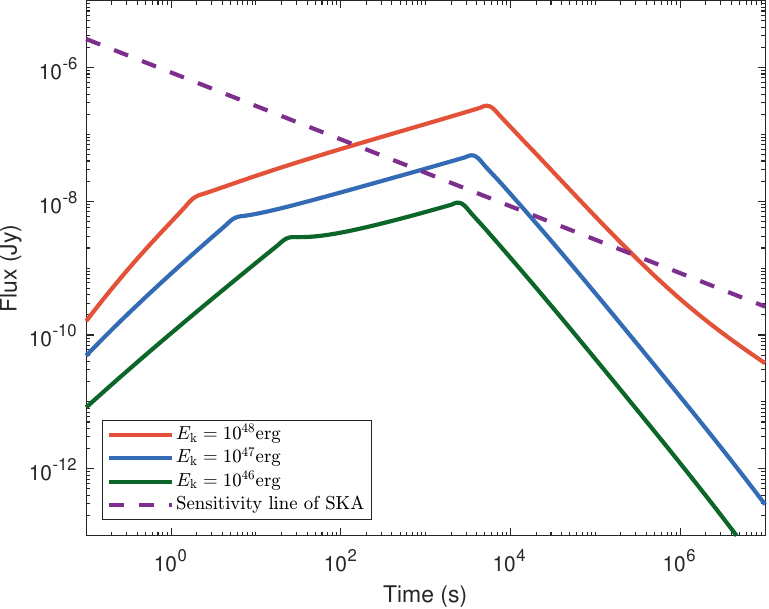}{0.3\textwidth}{(a)}
		\fig{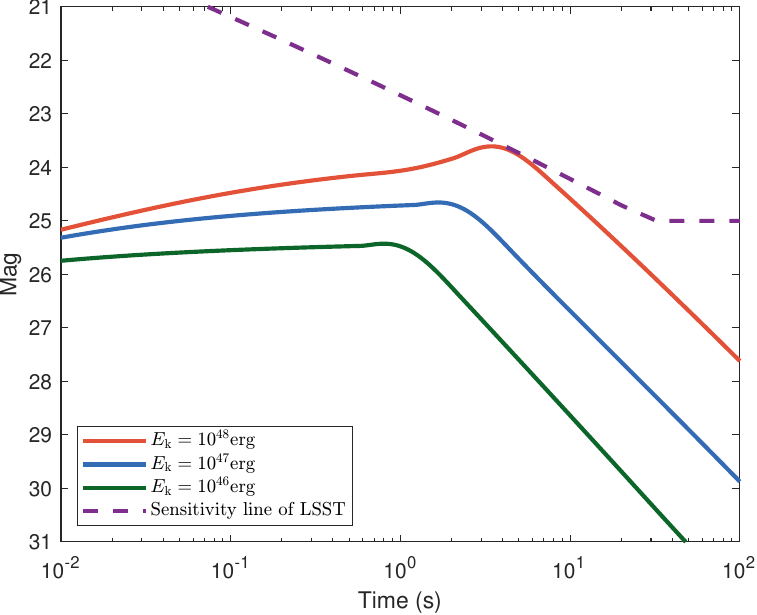}{0.29\textwidth}{(b)}
		\fig{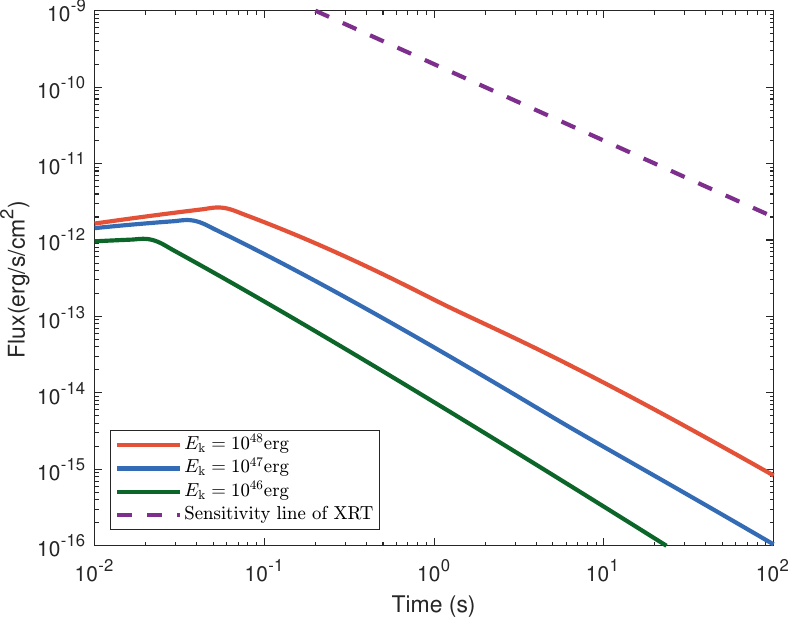}{0.3\textwidth}{(c)}}
	\caption{{Multi-wavelength afterglow light curves of FRB~20201124A for different ejecta kinetic energies: 
			$E_{\rm k} = 10^{46}~\mathrm{erg}$ (green), $10^{47}~\mathrm{erg}$ (blue), and $10^{48}~\mathrm{erg}$ (red). 
			Panels as in Figure~\ref{fig:1}.  The stellar-wind mass-loss rate is fixed at $\dot{M}_{\rm w} = 10^{-10} M_{\odot} \mathrm{yr}^{-1}$ in all three panels.}}
	\label{fig:3}
\end{figure*}

\begin{figure*}[ht]
	\gridline{\fig{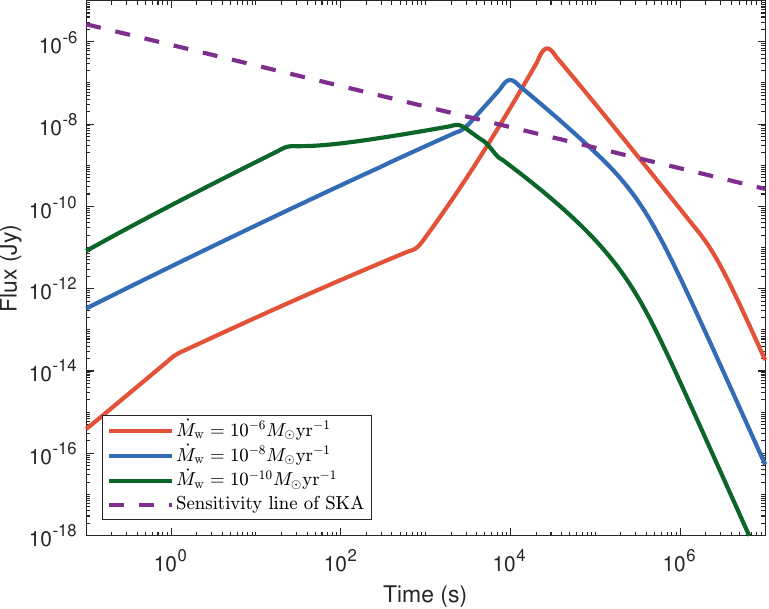}{0.3\textwidth}{(a)}
		\fig{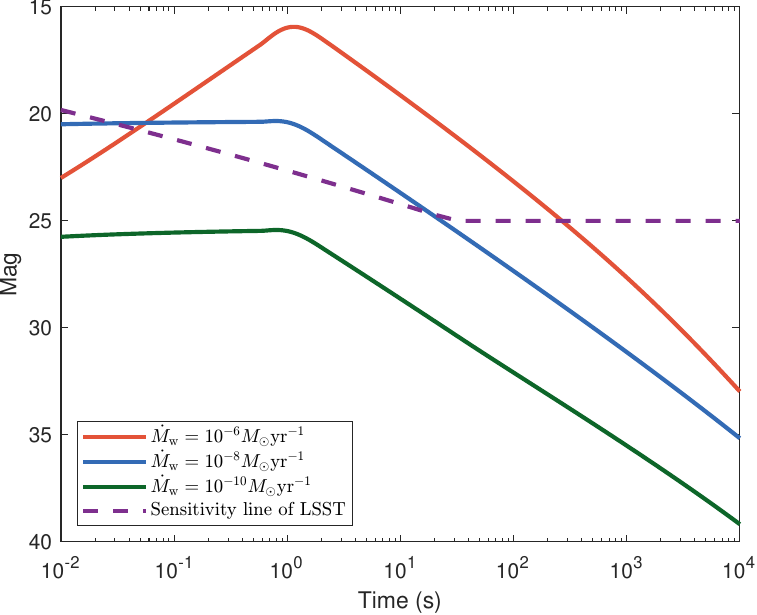}{0.29\textwidth}{(b)}
		\fig{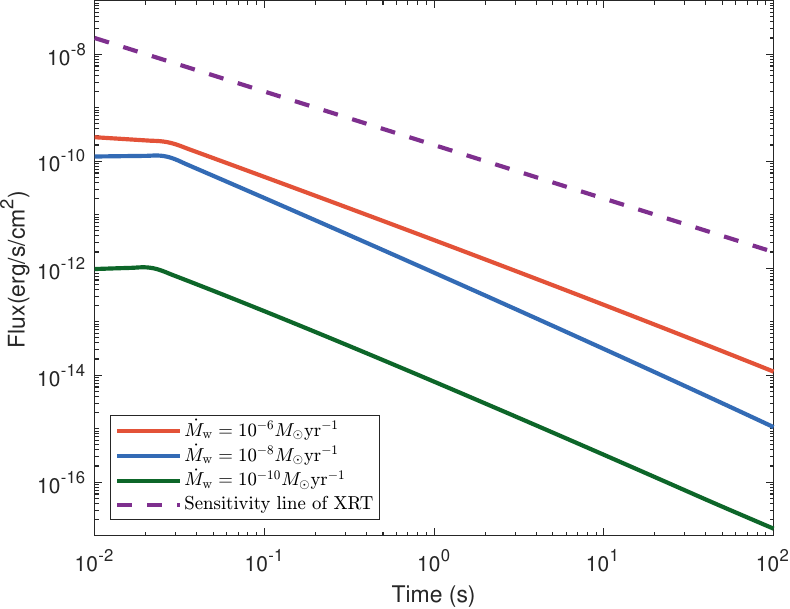}{0.3\textwidth}{(c)}}
	\caption{{Multi-wavelength afterglow light curves of FRB~20201124A for different stellar wind mass-loss rates: 
			$\dot{M}_{\rm w} = 10^{-10}~M_{\odot}~\mathrm{yr}^{-1}$ (green), $10^{-8}~M_{\odot}~\mathrm{yr}^{-1}$ (blue), and $10^{-6}~M_{\odot}~\mathrm{yr}^{-1}$ (red). 
			Panels as in Figure~\ref{fig:1}. The ejecta kinetic energy is fixed at $E_{\rm k} = 10^{46} \mathrm{erg}$ in all three panels.}}
	\label{fig:4}
\end{figure*}

\begin{figure*}[ht]
	\gridline{\fig{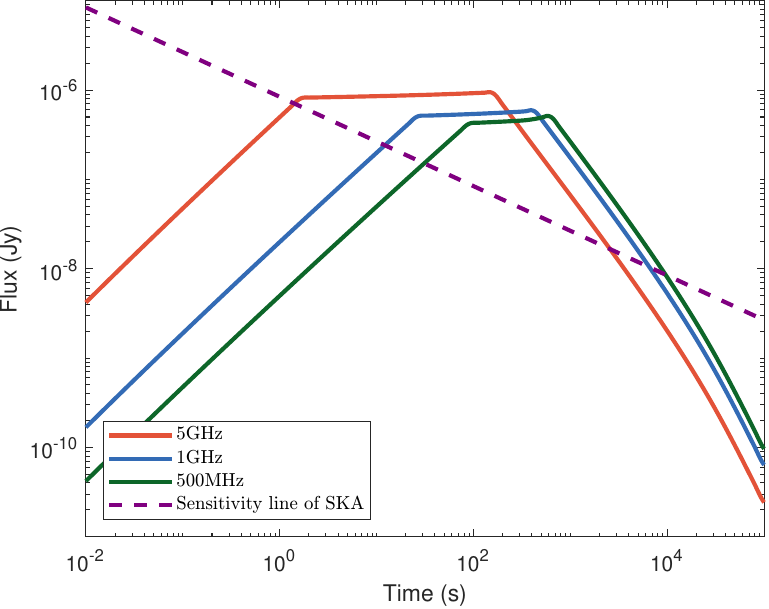}{0.3\textwidth}{(a)}
		\fig{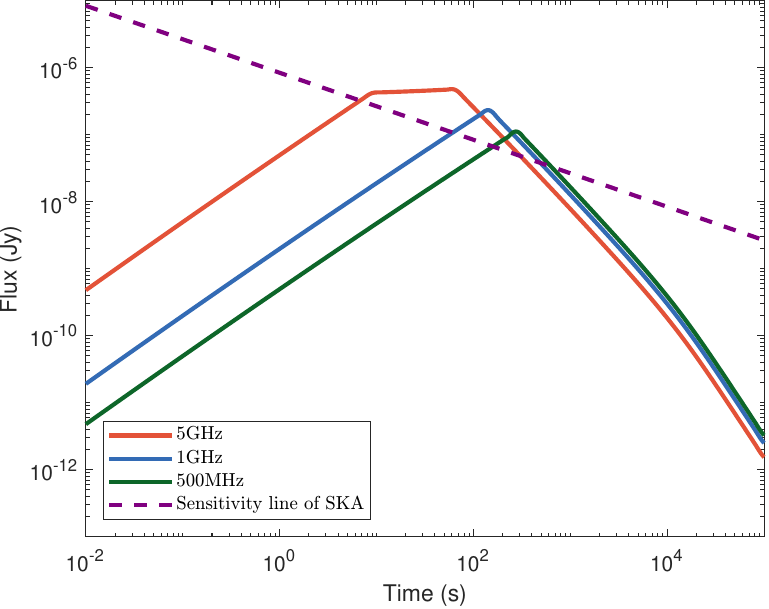}{0.3\textwidth}{(b)}
		\fig{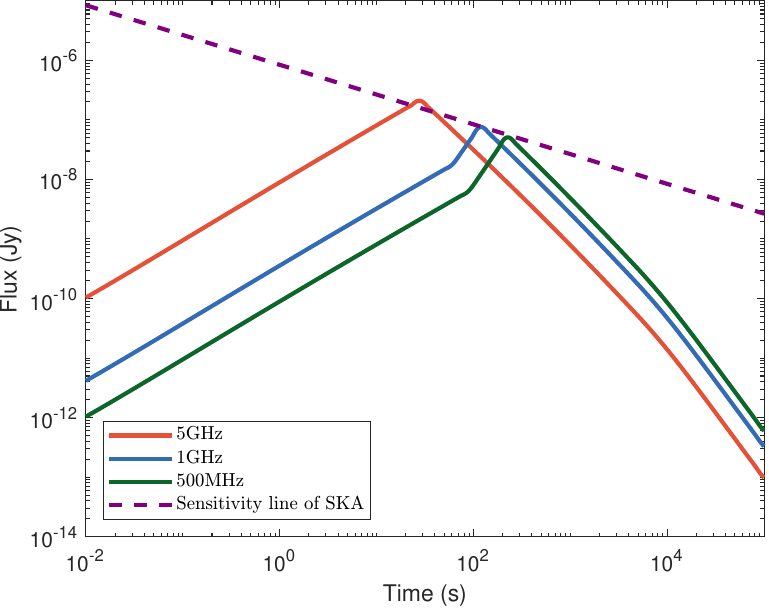}{0.3\textwidth}{(c)}}
	\gridline{\fig{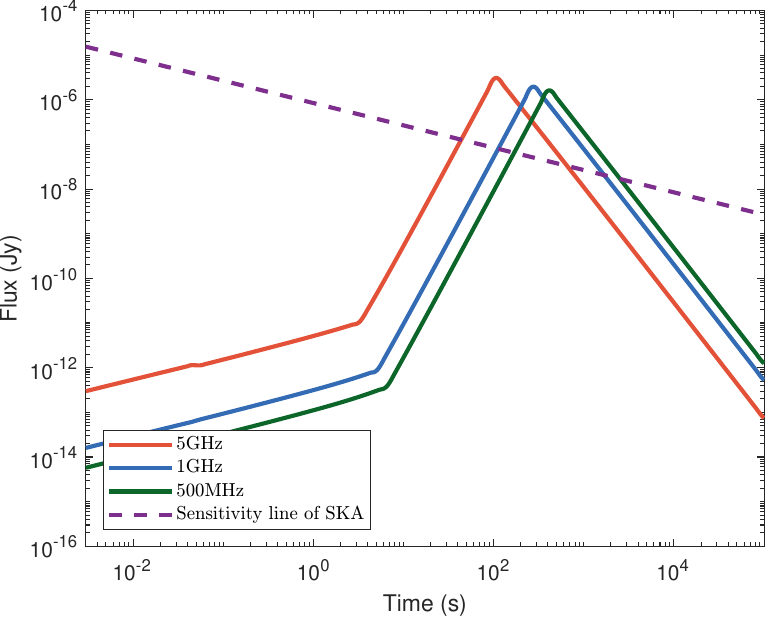}{0.3\textwidth}{(d)}
		\fig{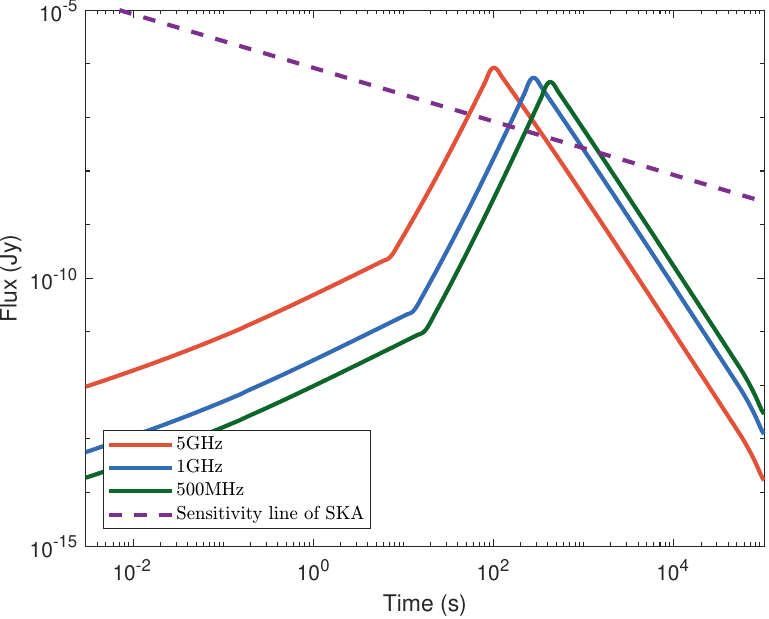}{0.3\textwidth}{(e)}
		\fig{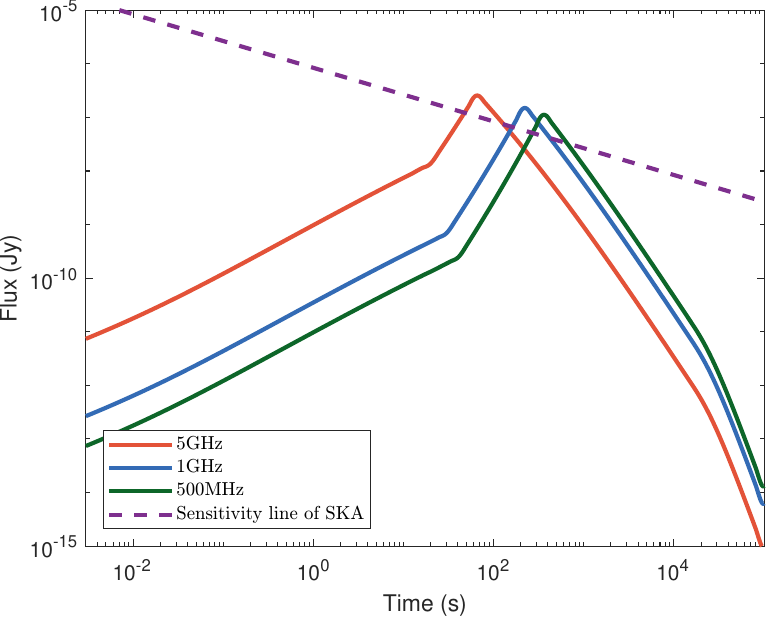}{0.3\textwidth}{(f)}}
	\caption{{Radio afterglow light curves of FRB~20200120E at different observing frequencies: 
			5~GHz (red), 1~GHz (blue), and 500~MHz (green). 
			Top row: different $E_{\rm k}$ values — (a) $10^{43}~\mathrm{erg}$, (b) $10^{42}~\mathrm{erg}$, (c) $10^{41}~\mathrm{erg}$, with fixed  $\dot{M}_{\rm w}$ value  of $10^{-11}~M_{\odot}~\mathrm{yr}^{-1}$. 
			Bottom row: different $\dot{M}_{\rm w}$ values — (d) $10^{-8}~M_{\odot}~\mathrm{yr}^{-1}$, (e) $10^{-9}~M_{\odot}~\mathrm{yr}^{-1}$, (f) $10^{-10}~M_{\odot}~\mathrm{yr}^{-1}$, with fixed $E_{\rm k}$  value of $10^{41}~\mathrm{erg}$.}}
	\label{fig:5}
\end{figure*}

\begin{figure*}[htbp]
	\gridline{\fig{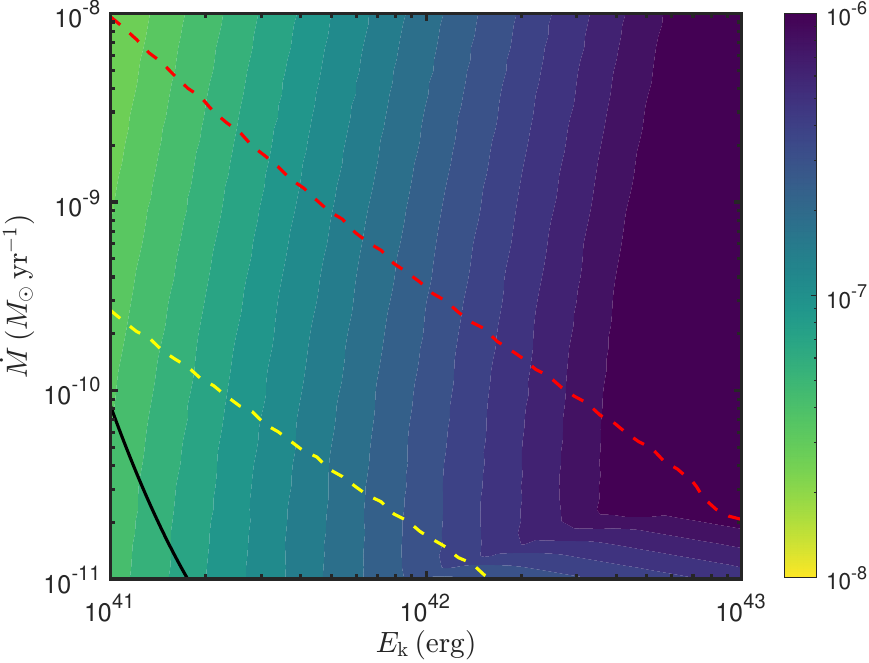}{0.5\textwidth}{(a)}
		\fig{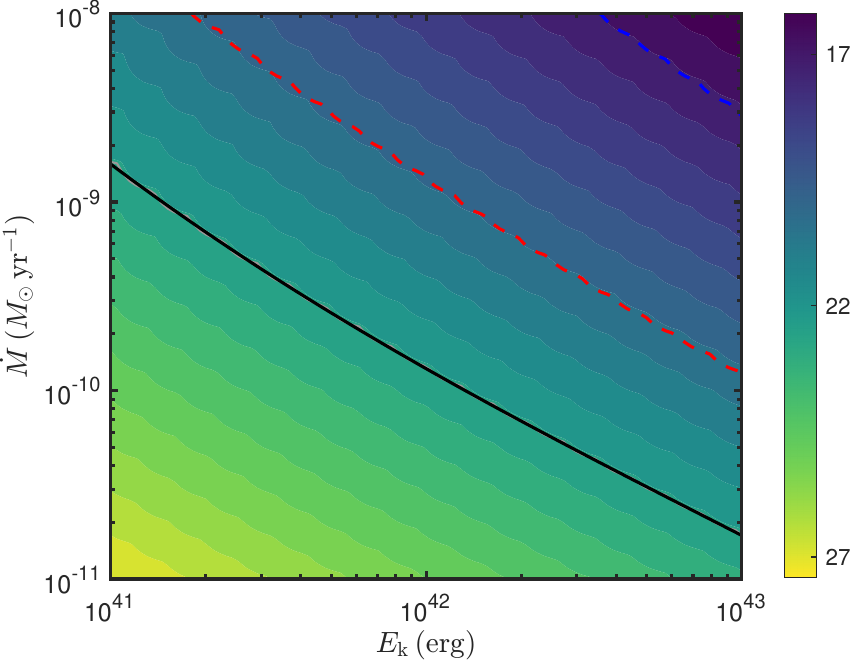}{0.5\textwidth}{(b)}}
	\caption{Contour maps illustrating the detectability of the afterglow from FRB~20200120E in the $(E_{\rm k}, \dot{M}_{\rm w})$ parameter space for an initial Lorentz factor $\eta = 200$. The left panel shows the results in the 1~GHz radio band, where the color contours represent the peak flux (in units of Jy) attained over the full light curve. The right panel corresponds to the optical $R$ band, with the color contours indicating the peak apparent magnitude. { In both panels, the black solid curve denotes the instrumental sensitivity threshold adopted in this work, corresponding to SKA at 1~GHz and LSST in the $R$ band, respectively. }  Dashed curves indicate the latest observing time at which the flux remains above the sensitivity limit.  In the 1~GHz band, the yellow and red dashed curves correspond to $10^{3}$~s and $10^{4}$~s, respectively, while in the $R$ band the red and blue dashed curves correspond to 10~s and 100~s, respectively. }
	\label{fig:9}
\end{figure*}

\begin{figure*}[ht]
	\gridline{\fig{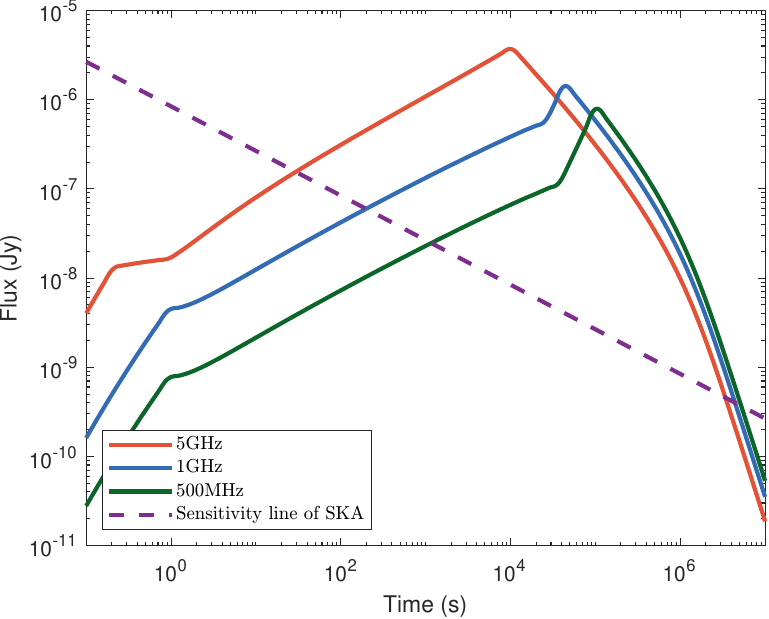}{0.3\textwidth}{(a)}
		\fig{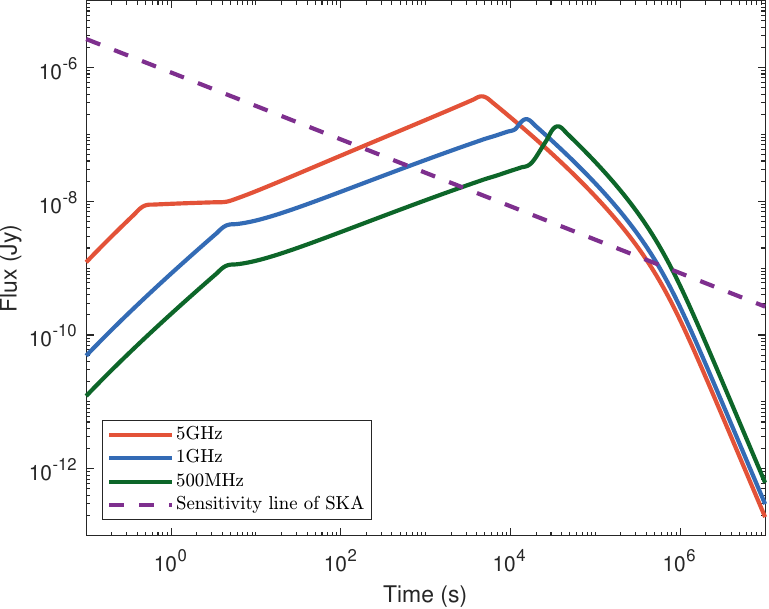}{0.3\textwidth}{(b)}
		\fig{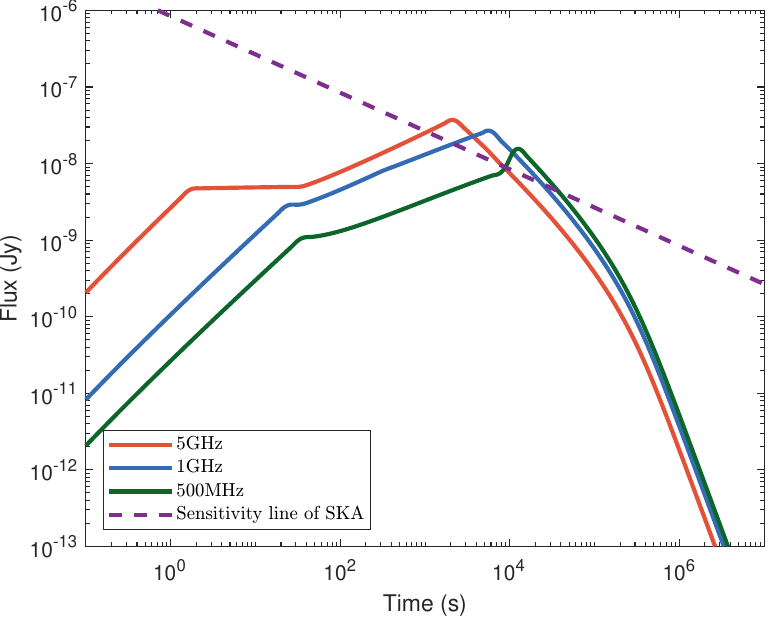}{0.3\textwidth}{(c)}}
	\gridline{\fig{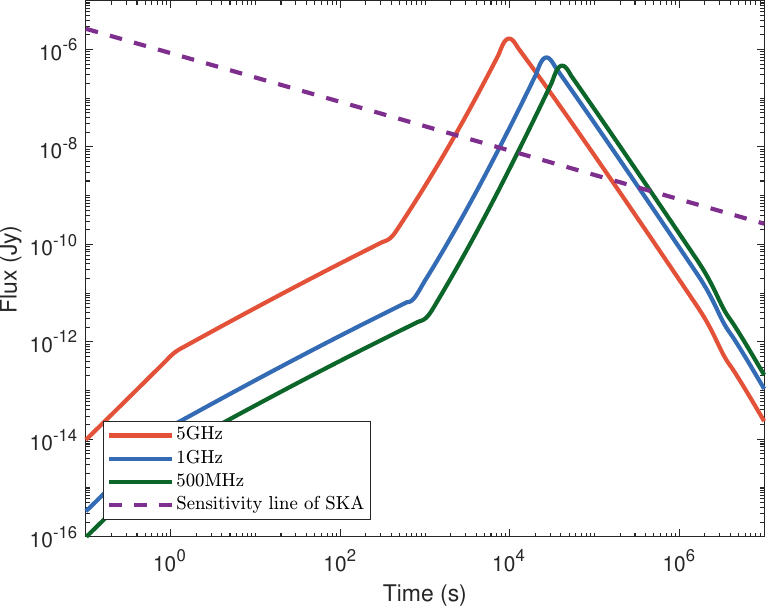}{0.3\textwidth}{(d)}
		\fig{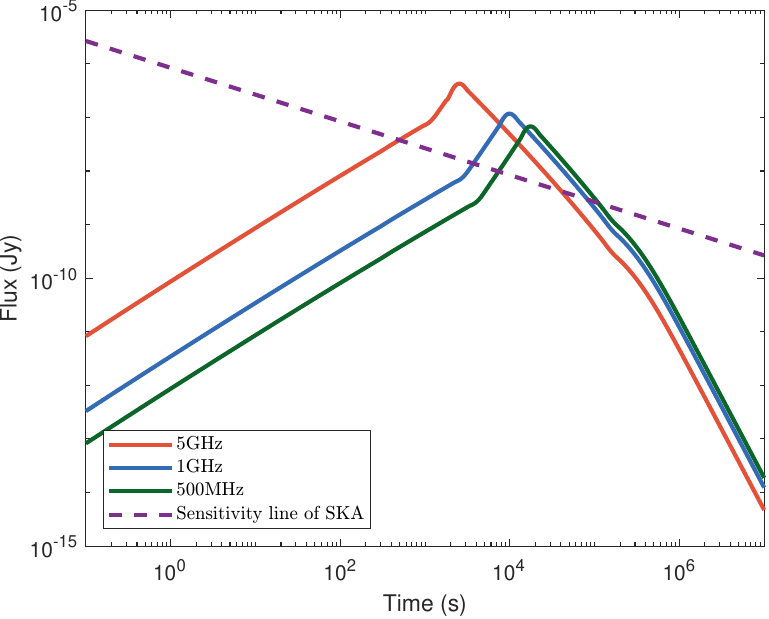}{0.3\textwidth}{(e)}
		\fig{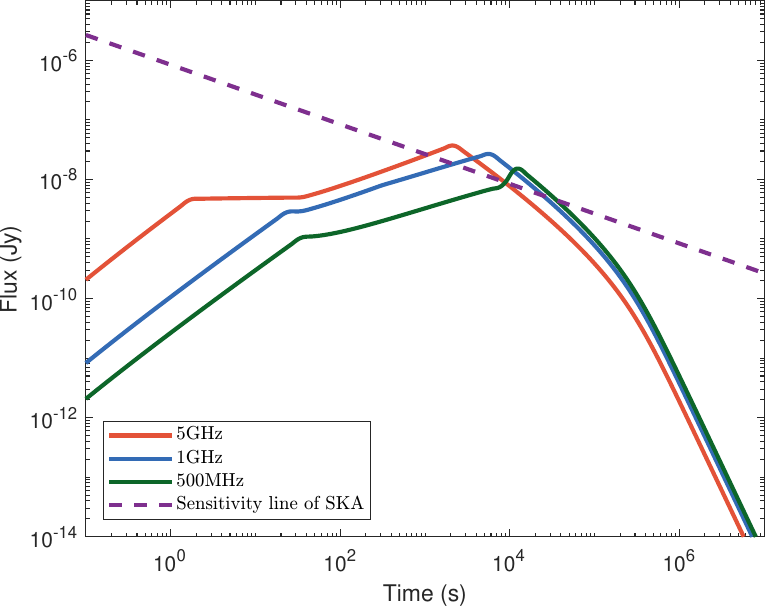}{0.3\textwidth}{(f)}}
	\caption{{Radio afterglow light curves of FRB~20201124A at different observing frequencies: 
			5~GHz (red), 1~GHz (blue), and 500~MHz (green). 
			Top row: different $E_{\rm k}$ values — (a) $10^{48}~\mathrm{erg}$, (b) $10^{47}~\mathrm{erg}$, (c) $10^{46}~\mathrm{erg}$,  with fixed  $\dot{M}_{\rm w}$ value  of $10^{-10}~M_{\odot}~\mathrm{yr}^{-1}$. 
			Bottom row: different $\dot{M}_{\rm w}$ values — (d) $10^{-6}~M_{\odot}~\mathrm{yr}^{-1}$, (e) $10^{-8}~M_{\odot}~\mathrm{yr}^{-1}$, (f) $10^{-10}~M_{\odot}~\mathrm{yr}^{-1}$, with fixed $E_{\rm k}$  value of $10^{46}~\mathrm{erg}$.}}
	\label{fig:6}
\end{figure*}

\begin{figure*}[htbp]
	\gridline{\fig{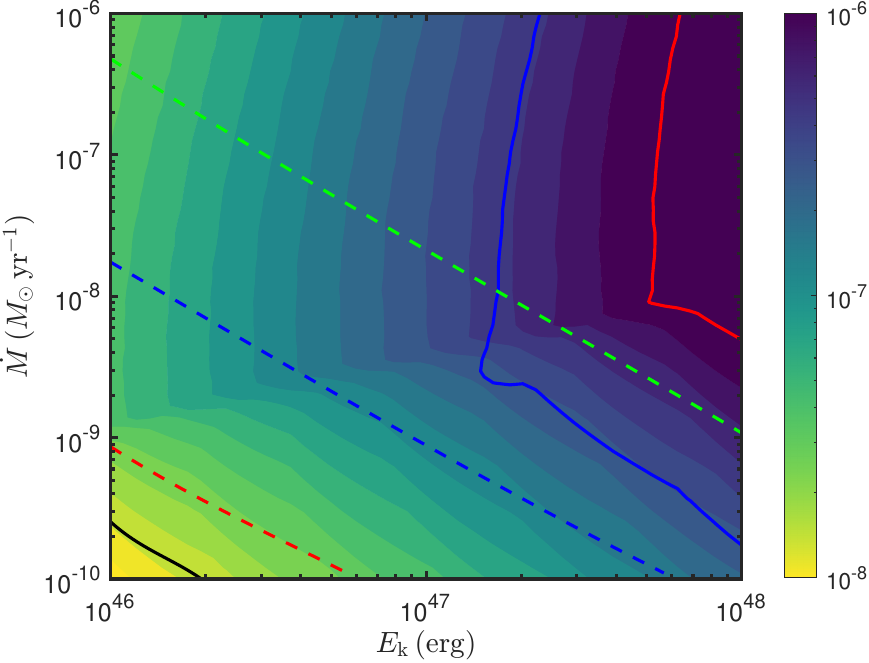}{0.5\textwidth}{(a)}
		\fig{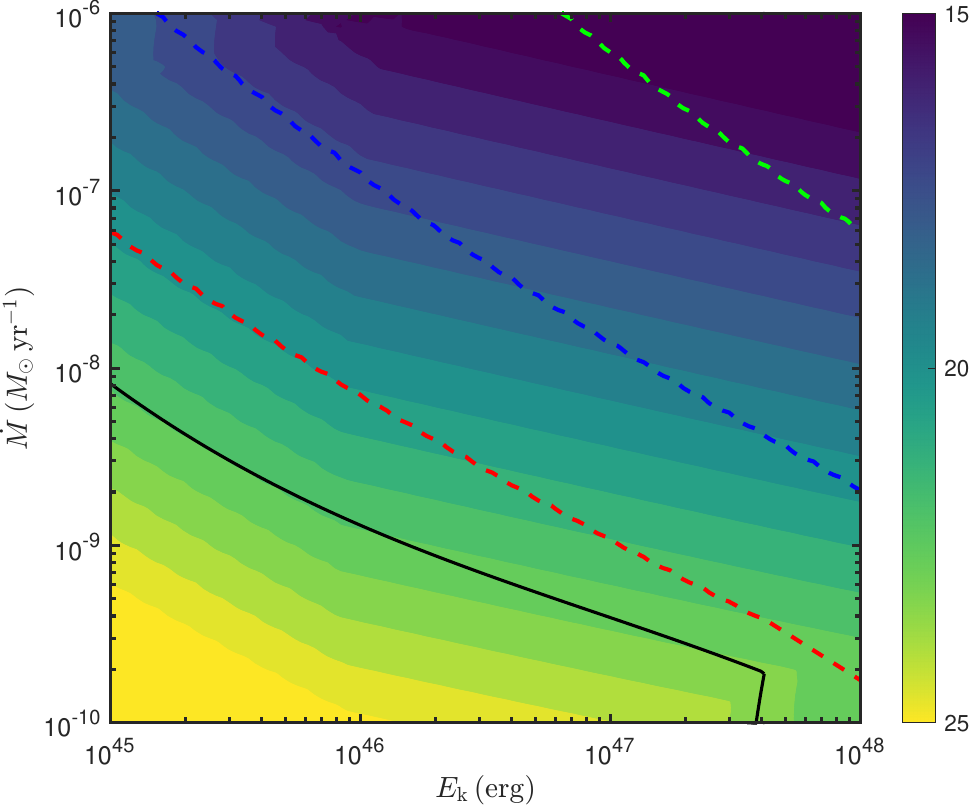}{0.45\textwidth}{(b)}}
	\caption{{Same as Figure \ref{fig:9}, but for FRB~20201124A. The left and right panels show the 1~GHz radio band and the optical $R$ band,respectively. In the 1~GHz band, the  red, blue, and green dashed lines correspond to $10^{4}$~s, $10^{5}$~s, and $10^{6}$~s, respectively.  And the red and blue solid lines indicate the parameter boundaries above which the predicted radio afterglow peak flux exceeds the level of the persistent radio emission at 1~GHz and 500~MHz, respectively.  In the $R$ band, the red, blue, and green dashed lines correspond to 10~s, 100~s, and 1000~s, respectively. }}
	\label{fig:10}
\end{figure*}

\begin{figure*}[htbp]
	\gridline{\fig{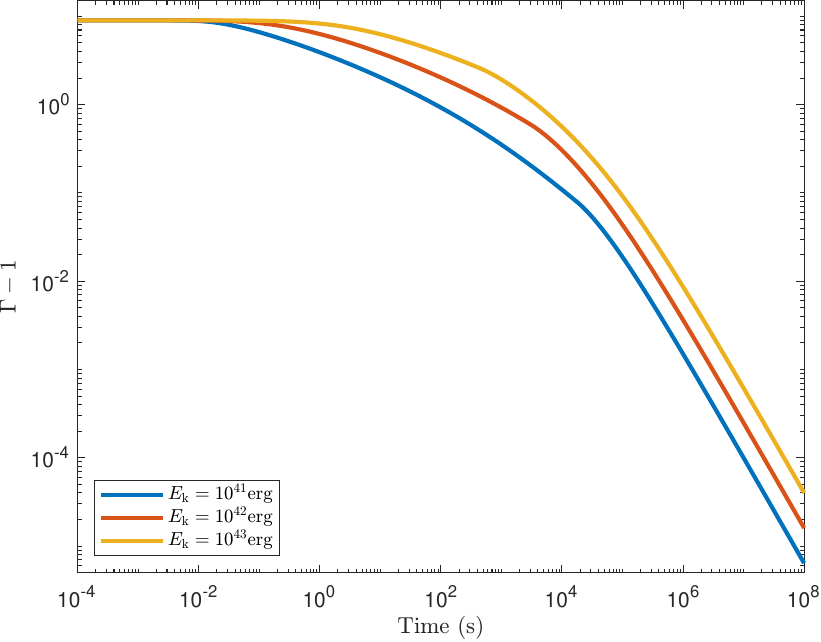}{0.5\textwidth}{(a)}
		\fig{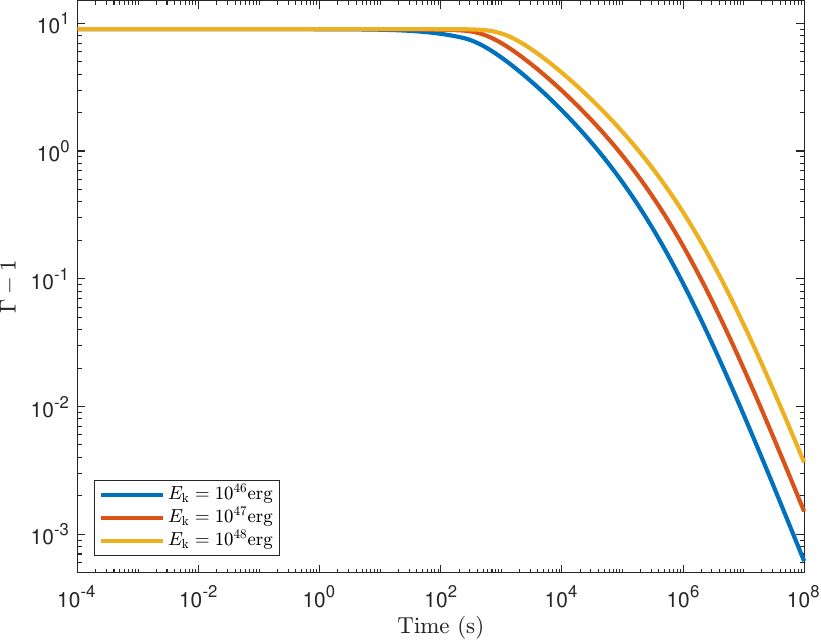}{0.5\textwidth}{(b)}}
	\caption{{Temporal evolution of the bulk Lorentz factor $\Gamma(t)-1$ for the two FRB cases. 
			Panel~(a): FRB~20200120E with $\dot{M}_{\rm w} = 10^{-11}~M_{\odot}~\mathrm{yr}^{-1}$ . 
			Panel~(b): FRB~20201124A with $\dot{M}_{\rm w} = 10^{-10}~M_{\odot}~\mathrm{yr}^{-1}$ .}}
	\label{fig:app6}
\end{figure*}

\begin{figure*}[htbp]
	\gridline{\fig{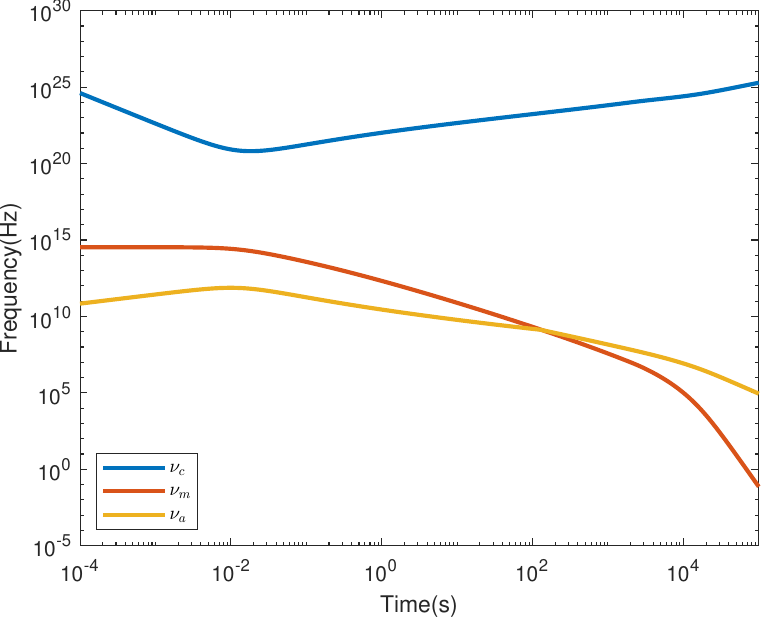}{0.5\textwidth}{(a)}
		\fig{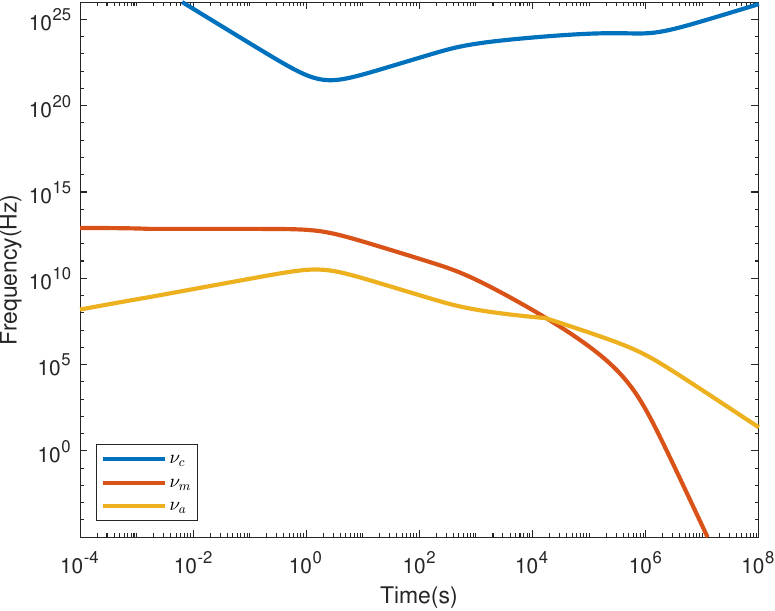}{0.5\textwidth}{(b)}}
	\caption{{Temporal evolution of the spectral break frequencies for the fiducial models. 
			Panel~(a): FRB~20200120E with $\dot{M}_{\rm w} = 10^{-11}~M_{\odot}~\mathrm{yr}^{-1}$ and $E_{\rm k} = 10^{42}~\mathrm{erg}$. 
			Panel~(b): FRB~20201124A with $\dot{M}_{\rm w} = 10^{-10}~M_{\odot}~\mathrm{yr}^{-1}$ and $E_{\rm k} = 10^{47}~\mathrm{erg}$. 
			}}
	\label{fig:app5}
\end{figure*}

\begin{figure*}[htbp]
	\gridline{\fig{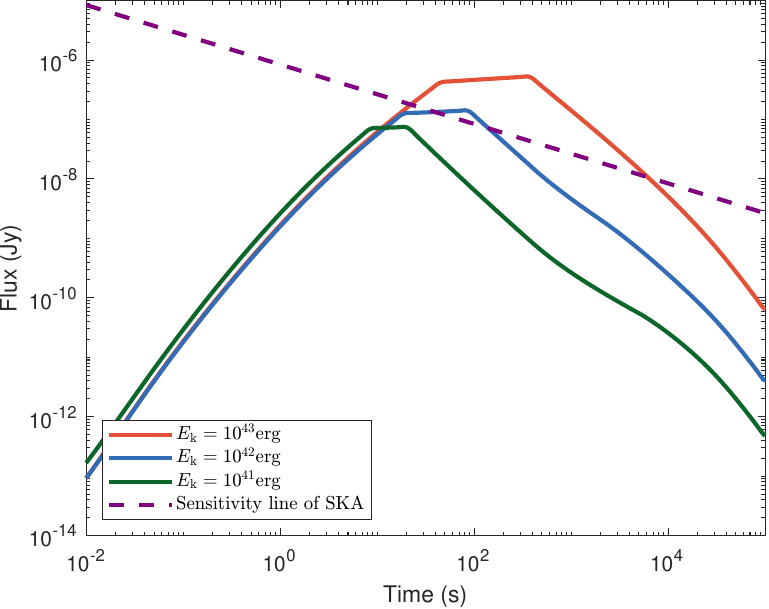}{0.3\textwidth}{(a)}
		\fig{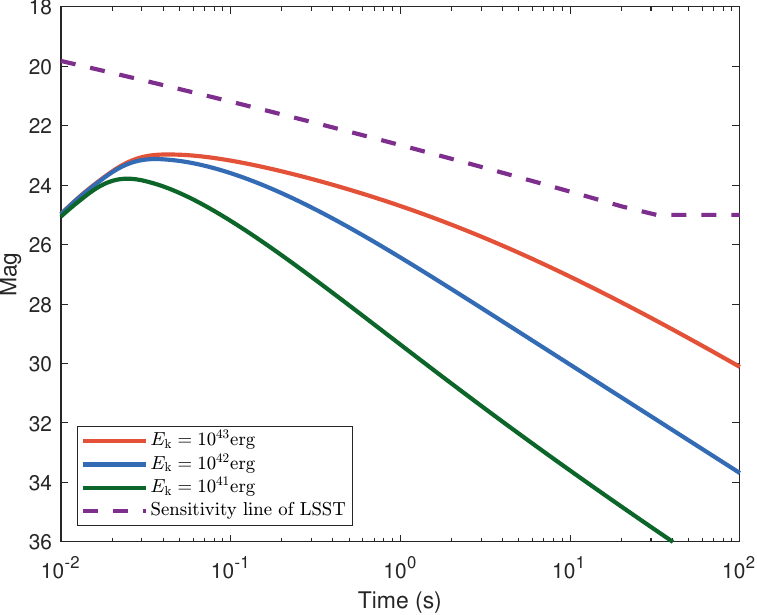}{0.29\textwidth}{(b)}
		\fig{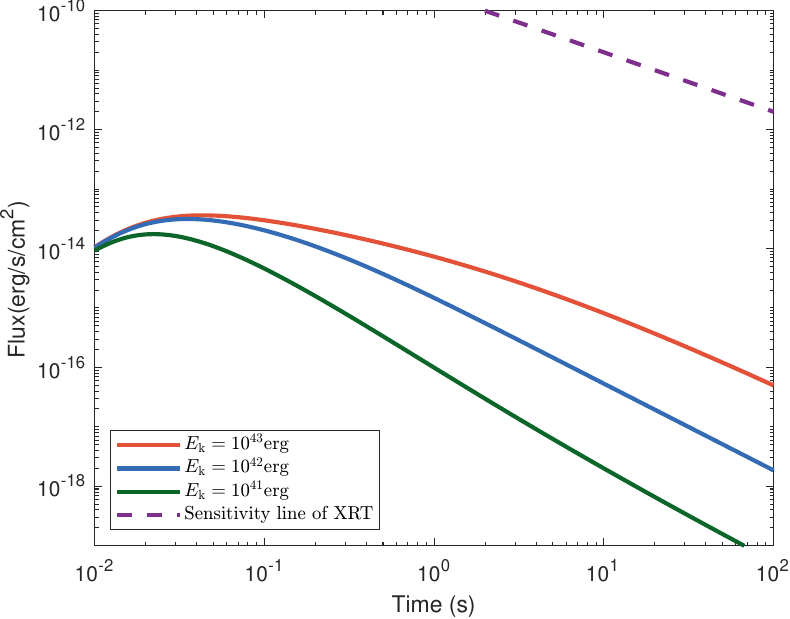}{0.3\textwidth}{(c)}}
	\caption{{Multi-wavelength afterglow light curves of FRB~20200120E for different ejecta kinetic energies: 
			$E_{\rm k} = 10^{41}~\mathrm{erg}$ (green), $10^{42}~\mathrm{erg}$ (blue), and $10^{43}~\mathrm{erg}$ (red).  The stellar-wind mass-loss rate is fixed at $\dot{M}_{\rm w} = 10^{-11} M_{\odot} \mathrm{yr}^{-1}$ in all three panels.
			Panel~(a): 1~GHz radio afterglow with SKA sensitivity limit (dotted purple). 
			Panel~(b): $R$-band optical afterglow with LSST sensitivity limit (dotted purple). 
			Panel~(c): 1~keV X-ray afterglow with Swift/XRT sensitivity limit (dotted purple).}}
	\label{fig:app1}
\end{figure*}

\begin{figure*}[htbp]
	\gridline{\fig{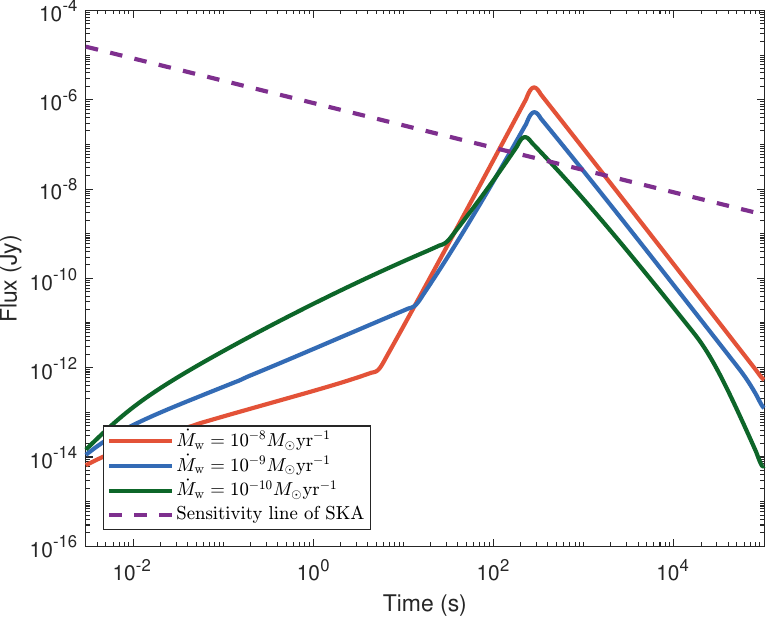}{0.3\textwidth}{(a)}
		\fig{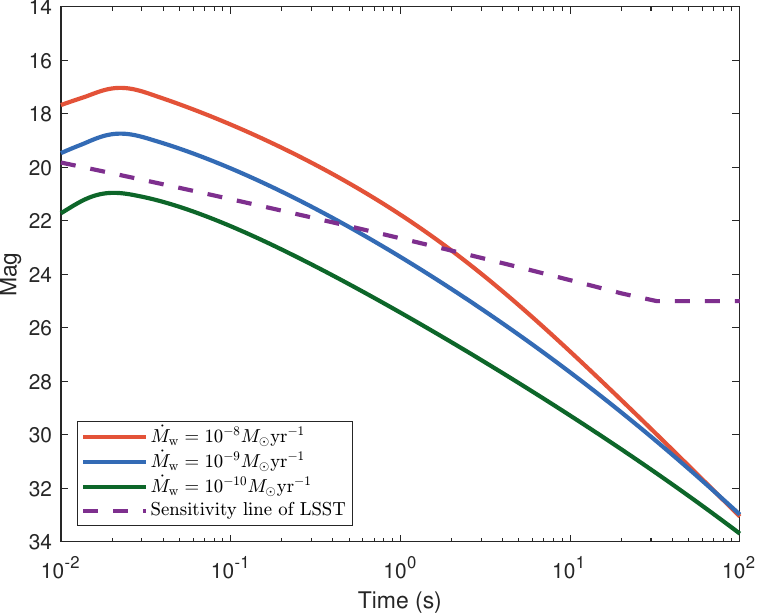}{0.29\textwidth}{(b)}
		\fig{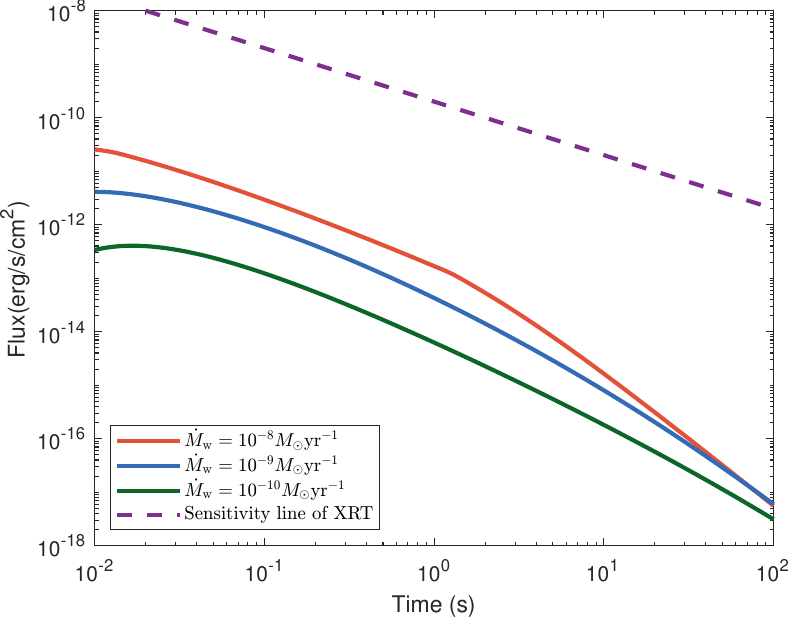}{0.3\textwidth}{(c)}}
	\caption{{Multi-wavelength afterglow light curves of FRB~20200120E for different stellar wind mass-loss rates: 
			$\dot{M}_{\rm w} = 10^{-10}~M_{\odot}~\mathrm{yr}^{-1}$ (green), $10^{-9}~M_{\odot}~\mathrm{yr}^{-1}$ (blue), and $10^{-8}~M_{\odot}~\mathrm{yr}^{-1}$ (red). 
			Panels as in Figure~\ref{fig:app1}. The ejecta kinetic energy is fixed at $E_{\rm k} = 10^{41} \mathrm{erg}$ in all three panels.}}
	\label{fig:app2}
\end{figure*}

\begin{figure*}[htbp]
	\gridline{\fig{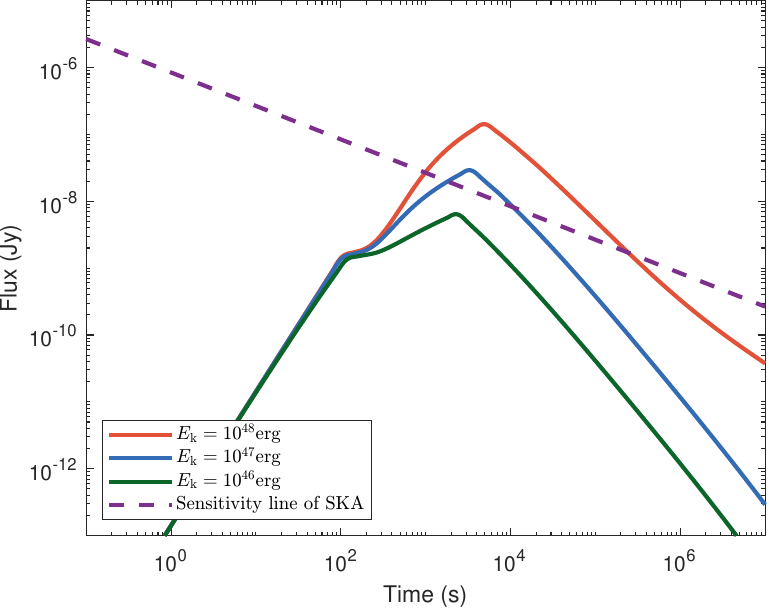}{0.3\textwidth}{(a)}
		\fig{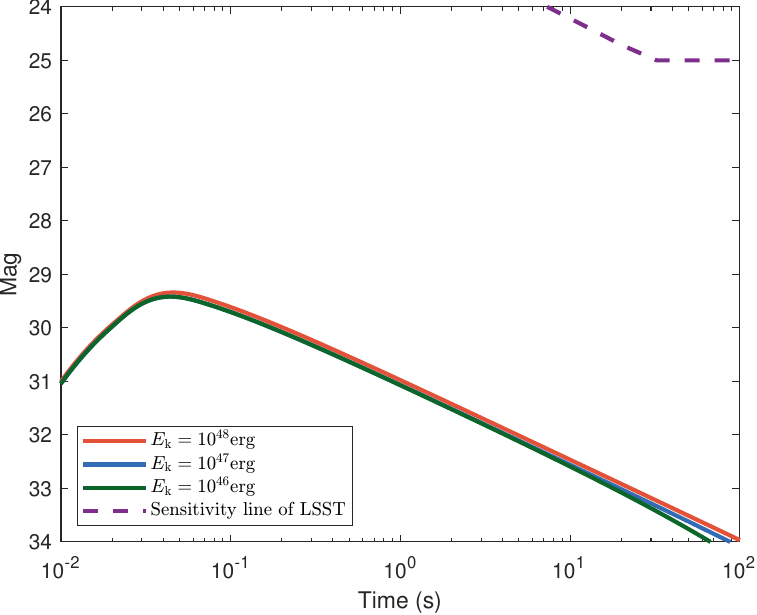}{0.29\textwidth}{(b)}
		\fig{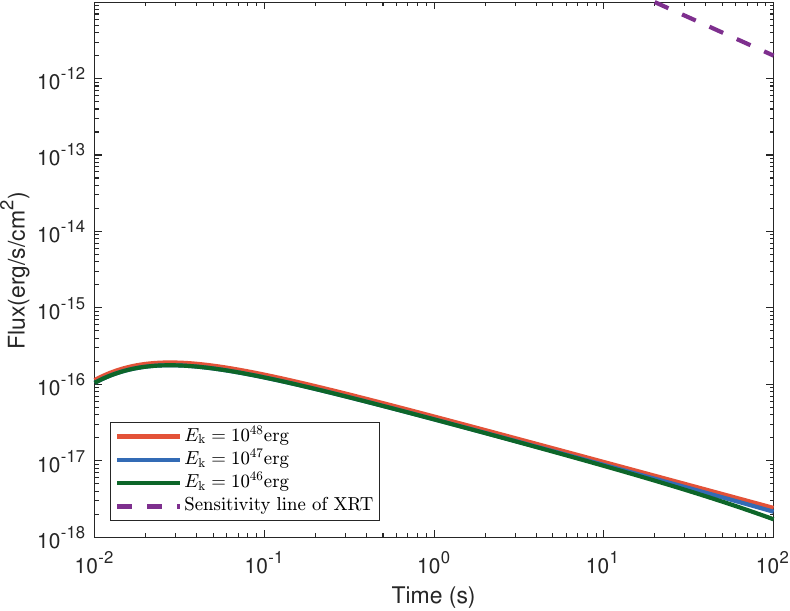}{0.3\textwidth}{(c)}}
	\caption{{Multi-wavelength afterglow light curves of FRB~20201124A for different ejecta kinetic energies: 
			$E_{\rm k} = 10^{46}~\mathrm{erg}$ (green), $10^{47}~\mathrm{erg}$ (blue), and $10^{48}~\mathrm{erg}$ (red). 
			Panels as in Figure~\ref{fig:app1}. The stellar-wind mass-loss rate is fixed at $\dot{M}_{\rm w} = 10^{-10} M_{\odot} \mathrm{yr}^{-1}$ in all three panels.}}
	\label{fig:app3}
\end{figure*}

\begin{figure*}[htbp]
	\gridline{\fig{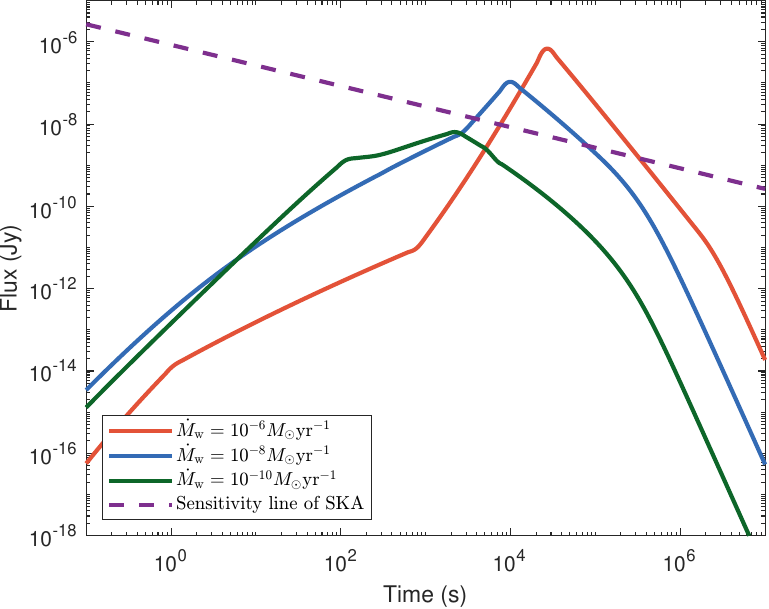}{0.3\textwidth}{(a)}
		\fig{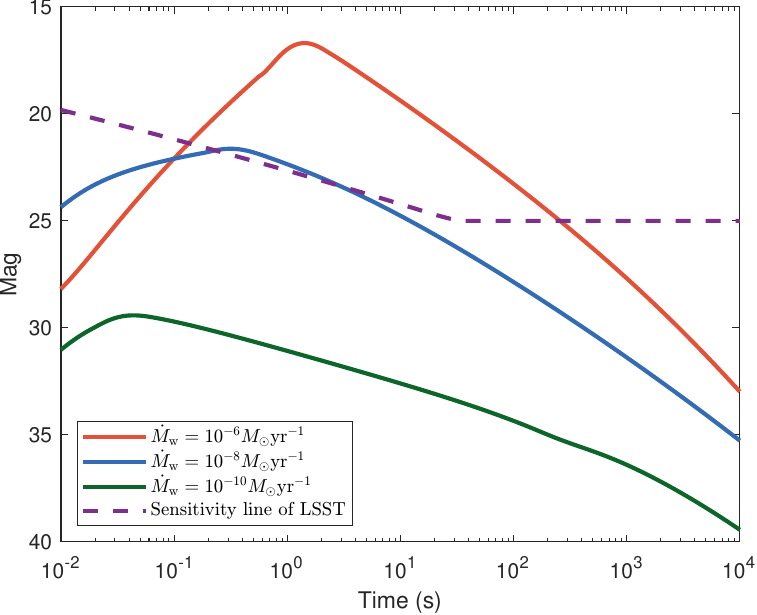}{0.29\textwidth}{(b)}
		\fig{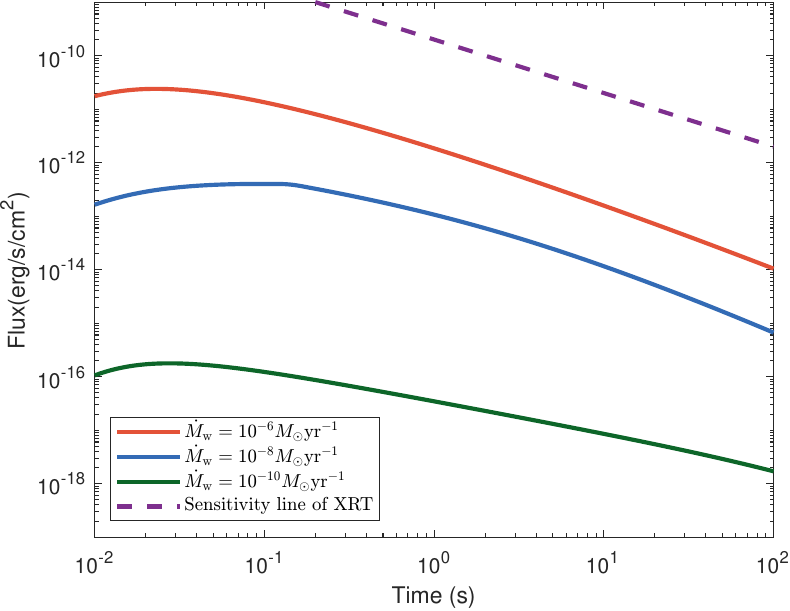}{0.3\textwidth}{(c)}}
	\caption{{Multi-wavelength afterglow light curves of FRB~20201124A for different stellar wind mass-loss rates: 
			$\dot{M}_{\rm w} = 10^{-10}~M_{\odot}~\mathrm{yr}^{-1}$ (green), $10^{-8}~M_{\odot}~\mathrm{yr}^{-1}$ (blue), and $10^{-6}~M_{\odot}~\mathrm{yr}^{-1}$ (red). 
			Panels as in Figure~\ref{fig:app1}. The ejecta kinetic energy is fixed at $E_{\rm k} = 10^{46} \mathrm{erg}$ in all three panels.}}
	\label{fig:app4}
\end{figure*}

\begin{figure*}[htbp]
	\gridline{\fig{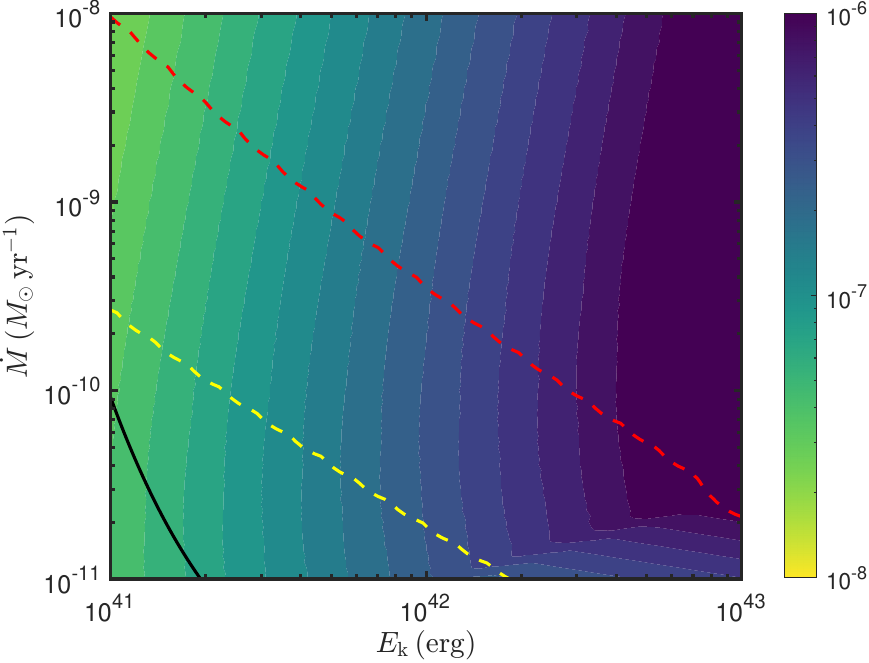}{0.5\textwidth}{(a)}
		\fig{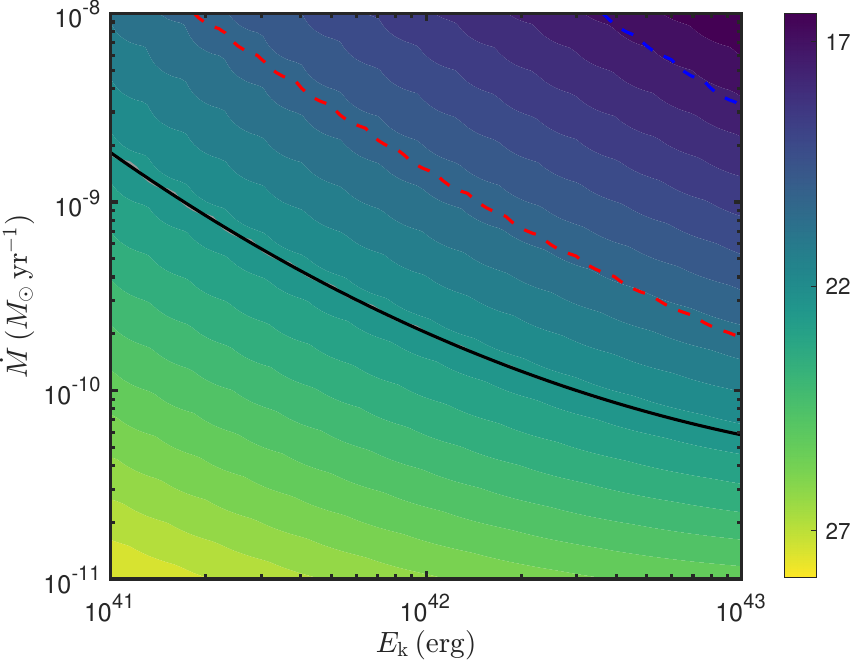}{0.5\textwidth}{(b)}}
	\caption{{Same as Figure~\ref{fig:9}, but for an initial Lorentz factor $\eta = 10$. The left panel shows the 1~GHz radio band, and the right panel corresponds to the optical $R$ band. Color contours indicate the peak flux (radio) or peak apparent magnitude (optical) over the full light curve. Solid curves mark the instrumental sensitivity thresholds, while dashed curves indicate the latest times at which the afterglow remains detectable. In the 1~GHz band, the yellow and red dashed curves correspond to $10^{3}$~s and $10^{4}$~s, respectively; in the $R$ band, the red and blue dashed curves correspond to 10~s and 100~s, respectively.  }}
	\label{fig:app7}
\end{figure*}

\begin{figure*}[htbp]
	\gridline{\fig{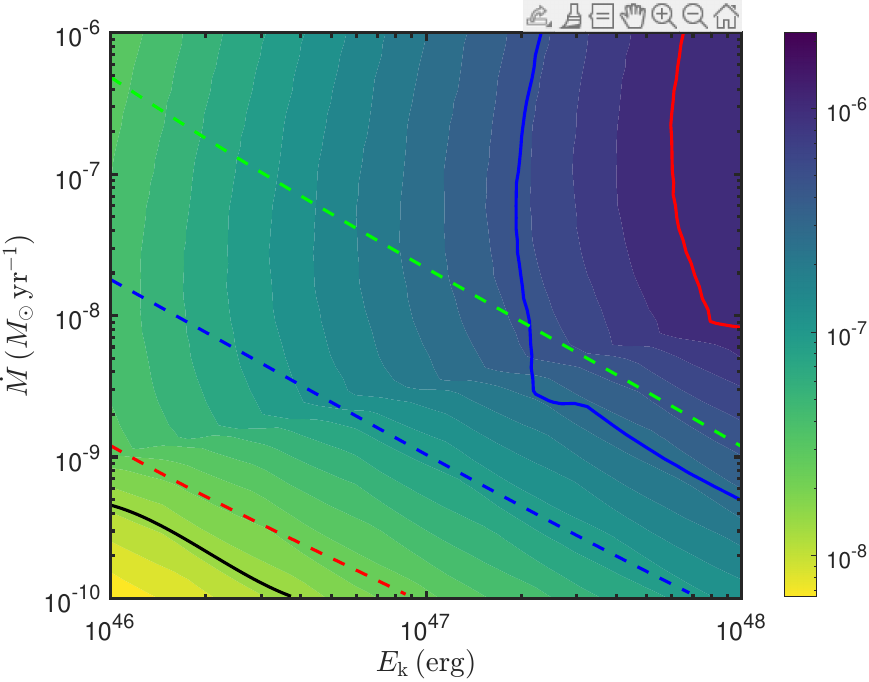}{0.5\textwidth}{(a)}
		\fig{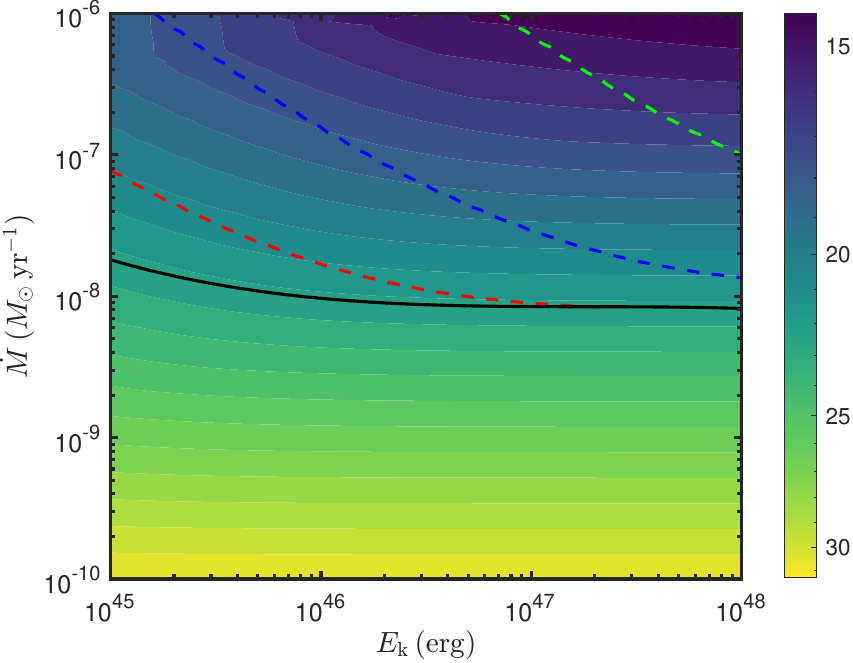}{0.5\textwidth}{(b)}}
	\caption{{Same as Figure \ref{fig:app7}, but for FRB~20201124A. The left and right panels show the 1~GHz radio band and the optical $R$ band,respectively. In the 1~GHz band, the  red, blue, and green dashed lines correspond to $10^{4}$~s, $10^{5}$~s, and $10^{6}$~s, respectively.  And the red and blue solid lines indicate the parameter boundaries above which the predicted radio afterglow peak flux exceeds the level of the persistent radio emission at 1~GHz and 500~MHz, respectively.  In the $R$ band, the red, blue, and green dashed lines correspond to 10~s, 100~s, and 1000~s, respectively. }}
	\label{fig:app8}
\end{figure*}

\clearpage
\appendix
{To assess how the initial bulk motion impacts our predictions, we recompute the multi-band light curves by adopting a lower initial Lorentz factor, $\eta=10$, while keeping all other parameters unchanged. The resulting evolutions of $\Gamma(t)$ and of the spectral break frequencies are shown in Figures~\ref{fig:app6} and \ref{fig:app5}. Figures~\ref{fig:app1}--\ref{fig:app8} reproduce the formats of Figures~\ref{fig:1}--\ref{fig:10} in the main text but with $\eta=10$.}

{For {FRB~20200120E}, lowering $\eta$ has only a minor effect on the radio band: the overall morphology and flux levels at 1\,GHz are nearly unchanged. The optical and X-ray bands are modestly dimmer, but the trends with $E_{\rm k}$ and $\dot{M}_{\rm w}$ remain similar to the fiducial results.}

{For {FRB~20201124A}, the response to a lower $\eta$ is markedly different. The radio light curves remain comparatively insensitive to $\eta$, but the optical and X-ray afterglows are \emph{strongly} suppressed when the stellar-wind mass-loss rate is low. In this low-density regime, the fluxes drop by large factors and become only weakly dependent on the ejecta kinetic energy $E_{\rm k}$. By contrast, at high $\dot{M}_{\rm w}$ the differences relative to the fiducial $\eta$ are much smaller, and the qualitative behaviors of the light curves are retained.}

{In summary, our main conclusions about radio detectability are robust against reasonable variations of the initial Lorentz factor. However, for FRB~20201124A the optical and X-ray prospects depend sensitively on $\eta$ in low-density winds, where lower $\eta$ renders these bands exceedingly faint.}

\end{document}